\documentclass[usenatbib]{mnras}

\usepackage[T1]{fontenc}
\usepackage{ae,aecompl}
\usepackage{graphicx}
\usepackage{subfig}
\usepackage{hyperref}
\usepackage{amsmath}
\usepackage{amssymb}
\usepackage{mathtools}
\usepackage{bm}
\usepackage{cleveref}

\title[Vortex-mediated mutual friction]{
Superfluid vortex-mediated mutual friction in non-homogeneous neutron star interiors
}

\author[M.Antonelli and B.Haskell]{
    M.~Antonelli 
    \thanks{E-mail: mantonelli@camk.edu.pl}, 
    B.~Haskell 
    \thanks{E-mail: bhaskell@camk.edu.pl}\\ 
    \\
    Nicolaus Copernicus Astronomical Center of the Polish Academy of Sciences, Bartycka 18, 00-716 Warszawa, Poland }

\begin{document}
    %
    %
    \pagerange{\pageref{firstpage}--\pageref{lastpage}} \pubyear{2019}
    \maketitle
    \label{firstpage}

\begin{abstract}
Understanding the average motion of a multitude of superfluid vortices in the interior of a neutron star is a key ingredient for most theories of pulsar glitches. 
In this paper we propose a kinetic approach to compute the  mutual friction force that is responsible for the momentum exchange between the normal and superfluid components in a neutron star, where the mutual friction is extracted from a suitable average over the motion of many vortex lines.
As a first step towards a better modelling of the repinning and depinning processes of many vortex lines in a neutron star, we consider here only straight and non-interacting vortices: we adopt a minimal model for the dynamics of an ensemble of point vortices in two dimensions immersed in a non-homogeneous medium that acts as a pinning landscape. 
Since the degree of disorder in the inner crust or outer core of a neutron star is unknown, we compare the two possible scenarios of periodic and disordered pinscapes.
This approach allows to extract the mutual friction between the superfluid and the normal component in the star when, in addition to the usual Magnus and drag forces acting on vortex lines, also a pinning force is at work. The effect of disorder on the depinning transition is also discussed.
\end{abstract}



\section{Introduction}
\label{sec:intro}

In the standard description of glitches - sudden spin-ups observed in the otherwise steadily decreasing rotational frequency of a pulsar - the neutron star is assumed to be divided in two components  that can rotate with slightly different angular velocities { (\citealt{baym+1969}, see also the review of \citealt{haskellmelatos2015})}: a normal component, the rotational period of which can be tracked by observing the pulsar electromagnetic emission, and a superfluid one, which is threaded by $\sim 10^{18}$ quantised vortices with circulation 
$\kappa = 2\times 10^{-3}$cm$^2$/s {(\citealt{HaskellsedrakianReview2018} and \citealt{chamel_review2017JApA} for recent reviews)}. 
The possibility of pinning between vortices and impurities in the inner crust \citep{andersonitoh1975}, or with quantized flux-tubes in the outer core {(\citealt{Muslimov1985}, see also \citealt{alpar2017JApA})}, forces the superfluid to lag behind the normal component during the spin-down  and a superfluid current develops in the frame of the the normal component. 
Part of the angular momentum carried by this superfluid neutron current may be sporadically released: unpinned by a still uncertain trigger, vortices suddenly transfer this excess of angular momentum to the normal component, causing a glitch.
Regarding the glitch trigger, the reason why $\sim 10^{12}$ out of $\sim 10^{18}$ vortices simultaneously unpin during a giant glitch in the Vela remains elusive, and proposed mechanisms range from hydrodynamical instabilities \citep{Glampedakis2009PhRvL,khomenko2019PhRvD}, starquakes {\citep{ruderman1991ApJ-III,franco2000ApJ,akbal2018,giliberti2019PASA}} and criticality of the vortex configuration~{\citep{Anderson1982PMagA,melatos2008}}. 

In complete analogy with the two-fluid description of superfluid $^4$He, the key ingredient that allows for a transfer of angular momentum between the superfluid and normal components in a pulsar is the mutual friction force \citep{langlois98,AndSid06}, which is also an important damping mechanism for the large-scale oscillations of rapidly rotating and cold neutron stars~\citep{mendell1991,LindblomMendell2000,haskell2009MNRAS}.

Recent observations of glitches in the Crab \citep{shaw+2018} and the Vela \citep{palfreyman2018Nat}
pulsars have been used to obtain constraints on the strength of the mutual friction \citep{ashton2019NatAs,montoli2020A&A}, which can shed light on the microscopic dissipative channels occurring in the star \citep{alpar84rapid,Jones1991a,epstein_baym92,graber+2018} and, indirectly, on the internal region where the glitch originates {{\citep{haskell_letter_2018,sourie2020MNRAS,pizzocheroAA20}}}. 

Despite glitches may be a promising tool to probe the internal physics of neutron stars {\citep{ho+2015,pizzochero+2017,montoli_eos_2020,guerci2020MNRAS}}, a complete description of the glitch dynamics would rely on connecting the macroscopic hydrodynamic modelling of the neutron superfluid in the pulsar interior with the complex physics of many quantized vortex lines at the mesoscale, i.e. over length scales of the order of the average vortex separation \citep{melatos_gpe_2011,melatos_Unpinning,Haskellhop}.
A similar problem arises also for laboratory superfluids, in particular for $^4$He, where different  models are used to describe the superfluid at different scales, ranging from quantum methods to resolve the vortex core, to the Hall-Vinen-Bekarevich-Khalatnikov (HVBK) hydrodynamics at the macroscopic scale, where the velocity fields are averaged over a fluid element \citep{Qbook}.
This average procedure should be carried out over a  portion of fluid containing several vortex lines. At this scale, the vortex core is not resolved, so that the dynamics can be described in terms of the vortex filament model pionereed by \cite{schwarz_85}. In particular, in a homogeneous medium, the motion of such a vortex filament is determined by the forces acting upon the line: the classical Magnus force and a drag force \citep{sonin_book_2016}. Under the simplifying assumption of straight vortex lines, the vortex-mediated mutual friction entering in the HVBK equations can be obtained \citep{Hall1956_II,Bekarevich1961,sourieKutta2020}.

To date, most glitch studies build on a minimal two-fluid hydrodynamic model that  is formally equivalent to the HVBK equations  \citep{prix2004,HaskellsedrakianReview2018}. However, differently from the case of  $^4$He, the neutron superfluid in a neutron star is immersed into a non-homogeneous background. For example, in the inner crust the dripped neutrons coexist with complex structures of bounded nucleons, like a crystal lattice or various kinds of pasta phases \citep{chamel_review_crust,newton2013pasta}. In such an environment the usual derivation of the HVBK mutual friction should be extended to take into account for the additional effect of inhomogeneities which may act as sites for pinning and hinder the outward motion of vortices \citep{andersonitoh1975,Alpar84a,seveso+2016}. 
The aim of this work is to make a step in this direction by calculating the average response to an externally fixed superfluid current of an ensemble of vortices in an non-homogeneous background. This allows to extract the mutual friction for a given value of the velocity lag between the superfluid neutrons and the normal component. 
This is analogous to the strategy used to extract the Hall resistivity \citep[e.g.][]{wang_PhysRevLett,Vinokur1993,Mawatari1997PhRvB} or the depinning transition \citep[e.g. ][]{reich1999,Reichhardt2009PhRvL,fily2010PhRvB} in type-II superconductors. 
In this sense, the approach resembles that of kinetic theory, which can be used to extract macroscopic transport properties (in this case the mutual friction) of out-of-equilibrium substances starting from the average properties of an ensemble of particles (in our case, point vortices).

The paper is organized as follows. In Sec 2 the hydrodynamic two-fluid formalism usually employed in neutron star applications is outlined and the macroscopic mutual friction is linked to an ensemble average of the local velocity of many vortex lines. 
In Sec 3 we assume a particular ``minimal'' model for vortex dynamics where non-interacting vortex segments are immersed into a non-homogeneous pinning landscape (a more general case is described in App \ref{app_solution}). 
Since the degree of impurities and disorder in a neutron star is uncertain {{(\cite{jones_amorph_99,sauls2020arXiv}, see also the review of \citealt{chamel_review_crust})}} two models for the pinning landscape are considered: one in which the pinning force field is periodic and one with quenched randomness (Sec 4).
In Sec 5 we present a numerical estimate of the average dynamics defined in Sec 3 for several sets of parameters. Sec 6 is devoted to study the behavior of the average vortex velocity in the adiabatic limit of very slow lag variations; the results relevant for pulsar glitch modelling are discussed. Our findings and conclusions are summarized in Sec 7.

\section{From vorticity transport to mutual friction}
\label{sec:mutual friction}

It is worth revisiting how to introduce a macroscopic friction force into a two-fluid HVBK-like model by using only the macroscopic fields of the hydrodynamic theory plus an auxiliary field that will be interpreted as the local average velocity of the quantized vortex lines in a fluid element \citep[see e.g. ][]{sonin_book_2016}. 

We adopt a simplified description for the matter in the inner crust of a neutron star, consisting of a charge neutral mixture of protons and electrons and superfluid neutrons.
The hydrodynamic equations of motion for this system reduce to a two-fluid model for two interacting species at zero temperature, the superfluid neutrons and a charge neutral normal fluid consisting of protons and electrons locked together by electromagnetic interactions on time scales shorter than those of interest for our problem \citep[see e.g.][]{langlois98}. 
Corrections to this model due to generation of heat via friction between the components is neglected.

We consider the Newtonian two-fluid equations derived by, for instance, \cite{prix2004} and \cite{andersson_comer2006CQG}. 
Each fluid satisfies an Euler-type equation for constituent ${\rm{x}},{\rm{y}}=n,p$ ($n$ for free neutrons, $p$ for the normal component)
\begin{equation}
    (\partial_t  + v_{\rm{x}}^j \nabla_j ) p_{\rm{x}}^i +\epsilon_{\rm{x}} v^{\rm{y x}}_j \nabla_i v_{\rm{x}}^j 
    + \nabla  \xi_{\rm{x}}  = F^i_{\rm{x}} \, ,
    \label{pizzobaldo}
\end{equation}
where  $\bm{v}_{\rm{y x}} = \bm{v}_{\rm{y }}-\bm{v}_{\rm{ x}}$ is the velocity of the $\rm{y}$-component as seen in the local rest frame of the $\rm{x}$-component and  
\begin{equation}
   p^{\rm{x}}_i =   v^{\rm{x}}_i + \epsilon_{\rm{x}} v^{{\rm{y x}}}_i 
   = (1-\epsilon_{\rm{x}}) v^{\rm{x}}_i + \epsilon_{\rm{x}} v^{{\rm{y}}}_i  
\end{equation}
is the momentum per unit mass. 
The non-dissipative entrainment coupling between the two fluids is encoded into the parameters $\epsilon_{\rm{x}}$ {{\citep[see e.g. the review of ][]{chamel_jltp_2017}}}.
The term $\xi_{\rm{x}}$ accounts for the chemical potential and gravity, while the mutual forces  per unit mass $F^i_{\rm{x}}$ on the RHS of \eqref{pizzobaldo} are (in principle unknown) phenomenological friction forces between the components that conserve the total momentum of the mixture (i.e. Newton's third law is implemented as $\rho_n F^i_{n}=-\rho_p F^i_{p}$, where $\rho_{\rm{x}}$ is the density of the x-constituent). 

To connect  $F^i_{\rm{x}}$ with the dynamics of quantized vortices at a smaller scale by using only the information contained into the LHS of \eqref{pizzobaldo}, we introduce an auxiliary macroscopic field $\bm{v}_L$ that is interpreted as the local coarse-grained velocity of the quantized vortex lines in a fluid element (see also \citealt{AndSid06}). 
A natural way of doing this is by considering the macroscopic vorticity  transport. Hence, let us rewrite \eqref{pizzobaldo} for the free neutrons as
\begin{equation}
\partial_t \bm{p}_n + L_{\bm{v}_n}  \bm{p}_n + \nabla \xi_n = \bm{F}_{n} \, ,  
\label{hvbkn1}
\end{equation}
where we used the Lie derivative of the momentum covector~\citep{andersson_turbulence}
\begin{equation}
 ( L_{\bm{v}_n} \bm{p}_n  )_i    =  v^j_n \nabla_j  p_{ni}  +  p^j_n \nabla_i  v_{nj}  \, .
\end{equation}
Writing \eqref{hvbkn1} in terms of the Lie derivative is convenient since 
\begin{equation}
L_{\bm{v}_n}  \bm{p}_n = \bm{\omega}_n \times \bm{v}_n + \nabla(\bm{p}_n \cdot \bm{v}_n) \, ,  
\label{lie2}
\end{equation}
where
\begin{equation}
\bm{\omega}_n = \nabla \times {\bm{p}}_{n} 
\label{hvbkn}
\end{equation}
is the macroscopic vorticity of the superfluid \citep{carter94,prix2004}. 
Taking the curl of \eqref{hvbkn1} and using \eqref{lie2} gives the macroscopic vorticity transport,
\begin{equation}
\partial_t \bm{\omega}_n + \nabla \times( \bm{\omega}_n \times \bm{v}_n )  
= \nabla \times \bm{F}_{n} \, . 
\label{hvbkn_curl}
\end{equation}
To introduce the average vortex velocity $\bm{v}_L$, we demand that vorticity is transported as 
\begin{equation}
\partial_t \bm{\omega}_n + \nabla \times( \bm{\omega}_n \times \bm{v}_L )  \, = 0  \, .
\label{hvbkn_curl2}
\end{equation}
Because of the cross product in the above equation, we define $\bm{v}_L$ to be orthogonal to $\bm{\omega}_n$ with no loss of generality (see \cite{gavassino2020MNRAS} for a complementary definition of this auxiliary velocity field in general relativity).
Now, equations \eqref{hvbkn_curl} and \eqref{hvbkn_curl2} are consistent only if 
\begin{equation}
\bm{F}_{n} = 
-\bm{\omega}_n \times(\bm{v}_L - \bm{v}_n) + \nabla \lambda_n 
\, .
\label{fmf_magnus}
\end{equation}
The constant of integration $\nabla \lambda_n$ will be neglected in the following, as it can be included into the term $\nabla \xi_n$ in \eqref{hvbkn1}.
Hence, modeling $\bm{v}_L$ is equivalent to providing a definite form to the mutual friction $\bm{F}_n$. In this vein, the quantized vortex lines are auxiliary entities that allow us to obtain $\bm{v}_L$ by means of an ensemble average in a local fluid element.

Extracting the average velocity of the vortex ensemble is trivial in the absence of quantum turbulence. 
In this case, the vortices in a fluid element are locally aligned along the unit vector $\hat{\boldsymbol{\kappa}}$ and the quantity $n_v= |\bm{\omega}_n|/ \kappa$ can be interpreted as the local density of vortex  lines per unit area of a surface orthogonal to $\hat{\boldsymbol{\kappa}}$, so that the coarse grained vorticity reads $\bm{\omega}_n = \hat{\boldsymbol{\kappa}} n_v$.
Furthermore, the average velocity $\bm{v}_L$ may be expressed as\footnote{
For notation convenience we assumed $\bm{v}_p \cdot \hat{\boldsymbol{\kappa}}=0$, see the Appendix \ref{app_solution} for the general case. The mutual friction $\bm{F}_{n}$ is left unchanged by this assumption.}
\begin{equation}
\bm{v}_L = \langle  \dot{\bm{x}} \rangle + \bm{v}_p 
= \frac{1}{N_v} \sum_{i=1}^{N_v}  \dot{\bm{x}}_i + \bm{v}_p \, ,  
\label{vL}
\end{equation}
where the sum is performed by considering a local fluid element containing a sample of $N_v$ quantized vortex lines with velocity $ \dot{\bm{x}}_i$ in the frame of the normal component.
Since the flow induced by the circulation around each of the $N_v$ vortices has zero average, the $\bm{v}_n$ appearing in \eqref{fmf_magnus} is the prescribed velocity field at infinity (i.e. the background prescribed flow in which the  fluid element is immersed). 
Therefore, 
\begin{equation}
\bm{F}_{n} = -  \kappa \, n_v\, \hat{\boldsymbol{\kappa}} \times 
( \langle \dot{\bm{x}} \rangle - \bm{v}_{np} ) = -  \kappa \, n_v\, \langle \bm{f}_M \rangle \, ,
\label{fmf_magnus2}
\end{equation}
where all the $\dot{\bm{x}}_i$ are calculated for a given and constant value of the velocity lag $\bm{v}_{np}$.
For later convenience we also introduced the {{local velocity 
$\langle \bm{f}_M \rangle = \hat{\boldsymbol{\kappa}} \times \langle \dot{\bm{x}} - \bm{v}_{np} \rangle $, 
that is an ensemble average of the (rescaled) Magnus lift force acting on a single vortex.
Since the force per unit mass $\bm{F}_n$ and the rescaled Magnus force $\langle \bm{f}_M \rangle $ are proportional,  in the following we will refer to both $\bm{F}_{n}$ and $\langle \bm{f}_M \rangle$ simply as ``mutual friction''.
}}

\section{Straight vortex dynamics in  a pinning landscape}
\label{sec:model}


We now provide a definite model for the dynamics of the single vortex velocity $\dot{\bm{x}}_i$ introduced in \eqref{vL}. 
We consider a straight segment of vortex line, modelled as a point with a vector $\boldsymbol{\kappa}$ attached, directed along the vortex itself.
The overdamped equation of motion for such a point vortex is \citep[see e.g.][]{sedrakianRepinning,ander_elastic_2011MNRAS}
\begin{equation}
\label{overdamped}
\rho_n \boldsymbol{\kappa} \times (\dot{\bm{x}}(t) - \bm{v}_{np} )
- \eta \, \dot{\bm{x}}(t) + \bm{\tilde{f}} 
\, = \, 0 \, ,
\end{equation}
where $\bm{x}(t)$ is the position of the vortex at time $t$ in a plane orthogonal to $\boldsymbol{\kappa}$. 
The additional force per unit length  $\bm{\tilde{f}}$ models the interaction with a non-homogeneous background medium in which the vortex is immersed  \citep{sedrakianRepinning,Haskellhop}.
To keep notation simple, we define $\dot{\bm{x}}$, $\bm{v}_{np}$ and $\bm{\tilde{f}} $  to be orthogonal to $ \hat{\boldsymbol{\kappa}}$ (otherwise we should explicitly include a projection on the plane orthogonal to $\hat{\boldsymbol{\kappa}}$).

Equation \eqref{overdamped} defines the dynamics of a vortex in the ensemble: the first term is the Magnus lift force and the next two terms are the longitudinal frictional force owing to the interaction of the vortex line with the constituents of the normal component.

In principle, the interaction with other vortex lines changes the value of $\bm{v}_{np}$, making it a dynamical variable that is function of the position of all vortex lines in the fluid element. In this work we will consider only the case in which  $\bm{v}_{np}$ is a constant value of the background velocity lag, namely we consider an ensemble of non-interacting point vortices.

To recast the differential equation \eqref{overdamped} into normal form, it is convenient to represent the vectors in the plane  as complex numbers, so that ($i$ is the imaginary unit)
\begin{equation}
i \,  (  \dot{x}  -  v_{np} )
- \mathcal{R} \, \dot{x} + f  
\, = \, 0 \, ,
\end{equation}
where $\mathcal{R}=\eta/\kappa \rho_n$ is a real parameter setting the the drag-to lift ratio and $f$ is the complex placeholder for the rescaled force $\bm{f} =\bm{\tilde{f}} /\kappa\rho_n$. 
It follows that
\begin{equation}
\label{langevinC}
\dot{x}(t) \, = \, \frac{f  - i v_{np}}{\mathcal{R}-i}\,
  \, = \,
 \cos{\theta_d} \,  e^{-i \theta_d} (v_{np}+if )
 \, ,
\end{equation}
where the dissipation angle $\theta_d \in [0,\pi/2) $ is related to the drag via $\mathcal{R}=\tan{\theta_d}$. When $\bm{f} =0$, the positive parameter $\theta_d$  reduces to the angle between the vortex velocity and the lag $\bm{v}_{np}$, see e.g. \cite{epstein_baym92} and \cite{celora2020MNRAS}.
The drag coefficient $\mathcal{R}$ and the dissipation angle $\theta_d$ are typically small,
so that, if pinning forces average to zero over the ensemble, the vortex lines mostly flow with a velocity close to the azimuthal superfluid velocity \citep{Jones1991a}.
Equation \eqref{langevinC} is equivalent to 
\begin{equation}
\label{normal}
\dot{\bm{x}} 
\, = \,  
\cos{\theta_d} \, R^{-1}_{\theta_d} \, 
( \bm{v}_{np} +  R_{\pi/2}   \bm{f}  ) 
 \, ,
\end{equation}
where the matrix $R_{\alpha}$ performs an anticlockwise rotation of an angle $\alpha$  in the plane.  
Alternatively, equation \eqref{normal} can be written by using a three-dimensional formalism
\begin{multline}
\label{xdotttt}
\dot{\bm{x}} 
\, = \,  
\frac{1}{1+\mathcal{R}^2} \, 
\left( \bot  \, \bm{v}_{np} +  \hat{\boldsymbol{\kappa}} \times   \bm{f}    \right)+
\\
+\frac{\mathcal{R}}{1+\mathcal{R}^2} \, 
\left(  \bot  \, \bm{f}  - \hat{\boldsymbol{\kappa}} \times    \bm{v}_{np}    \right) \, ,
\end{multline}
where $\bot_{ij} = \delta_{ij}-{\hat{\kappa}}_i {\hat{\kappa}}_j$ is the projector on the plane orthogonal to $\hat{\boldsymbol{\kappa}} $. 
Thanks to \eqref{fmf_magnus2}, the mutual friction reads
\begin{multline}
\label{mutualone}
\frac{\bm{F}_{n}}{|\bm{\omega}_n|} 
\, = \,  
\frac{\mathcal{R}^2}{1+\mathcal{R}^2} \, 
  \hat{\boldsymbol{\kappa}} \times    \bm{v}_{np}    
+
\frac{1}{1+\mathcal{R}^2} \, 
 \bot  \langle \bm{f}  \rangle -
 \\
 -
\frac{\mathcal{R}}{1+\mathcal{R}^2} \, 
\left( \hat{\boldsymbol{\kappa}} \times  \langle \bm{f}  \rangle
+ \bot \,  \bm{v}_{np}   \right)
  \, ,
\end{multline}
where the coarse grained field $\langle \bm{f} \rangle$ is still unknown and should be locally calculated as
\begin{equation}
    \langle \bm{f} \rangle = \frac{1}{N_v} \sum_{i=1}^{N_v} \bm{f}(\bm{x}_i(t))
\end{equation}
The usual  HVBK mutual friction \citep{donnelly_book}, that is linear in the lag $\bm{v}_{np}$, is recovered when $ \langle \bm{f}  \rangle = 0$: 
\begin{equation}
\label{mf_standard}
\frac{\bm{F}_{n}}{|\bm{\omega}_n|} 
\, = \,  
\frac{\mathcal{R}^2}{1+\mathcal{R}^2} \, 
  \hat{\boldsymbol{\kappa}} \times    \bm{v}_{np}    
+
\frac{\mathcal{R}}{1+\mathcal{R}^2} \, 
\hat{\boldsymbol{\kappa}} \times  (\hat{\boldsymbol{\kappa}} \times  \bm{v}_{np})  \, ,
\end{equation}
in accordance with \cite{AndSid06}. 
Clearly, the results \eqref{mutualone} and \eqref{mf_standard} are not the most general form of mutual friction one could obtain on the basis a purely geometric argument, but depend on the  particular equation of motion \eqref{overdamped} assumed for a vortex.
In Appendix A the mutual friction $\bm{F}_{n}$ is derived for a more general 
equation of vortex motion (cf \eqref{mutualone} with \eqref{fmf_tottta}).

\subsection{Non-linear mutual friction}
\label{sec_nonlinear}
 
Before moving to simulate the dynamics defined by \eqref{normal}, it can be useful to rewrite the mutual friction \eqref{mutualone} in a  way that resembles more closely the one used in many pulsar glitch models of the kind first proposed by \cite{Alpar84a}. 
According to this theory, the coupling between the superfluid and the normal crust is achieved via thermal creep of the vortex lines against pinning energy barriers, in analogy with flux creep in Type-II superconductors \citep{anderson_Kim_64}. 
Therefore, it is natural to try to encode the effect of pinning into the HVBK-like expression \eqref{mf_standard} in terms of unpinning probability, that is the weight function for the instantaneous number of (unpinned) moving vortices \citep[e.g. ][]{jahan2006ApJ,link2014ApJ}. 
In fact, unless an additional dependence of $\mathcal{R}$ on $\bm{v_{np}}$ or $\langle \dot{\bm{x}}\rangle$ is explicitly taken into account \citep{celora2020MNRAS}, the standard HVBK form \eqref{mf_standard} is linear in the lag $\bm{v}_{np}$ and this does not allow to describe the full sequence of dynamical phases expected in a glitch (unpinning, relaxation and repinning). 

First, it is convenient to introduce a physical right-handed orthonormal basis $(\hat{\bm{e}}_1,\hat{\bm{e}}_2)$ on the plane orthogonal to $\hat{\boldsymbol{\kappa}}$, defined by the unit vectors
\begin{equation}
\hat{\bm{e}}_1 = -\hat{\boldsymbol{\kappa}} \times \frac{\bm{v}_{np}}{|\bm{v}_{np}|} 
\qquad \qquad 
\hat{\bm{e}}_2 = \frac{\bm{v}_{np}}{|\bm{v}_{np}|} \, .
\label{basis}
\end{equation}
In pulsar glitch applications, where the motion of the superfluid along $\hat{\bm{e}}_2$ is assumed to be azimuthal, the unit vector $\hat{\bm{e}}_1$ would be directed radially outward. In the limit of ``free'' vortices (i.e. $\bm{f}=0$), the velocity of a single vortex $\dot{\bm{x}}$ in \eqref{xdotttt} is constant and coincides with the local average $\langle \dot{\bm{x}} \rangle $ and we have
\begin{equation}
 \bm{\dot{x}} = \bm{v}_{fr} = v^{fr}_1 \, \hat{\bm{e}}_1 + v^{fr}_2 \, \hat{\bm{e}}_2  \qquad (\text{for } \bm{f}=0)\, , 
\label{basis1}
\end{equation}
where 
\begin{equation}
 v^{fr}_1 = \sin\theta_d \cos\theta_d  \,  v_{np}
 \qquad  
 v^{fr}_2 = \cos^2\theta_d \,  v_{np} \, .
\label{basis2}
\end{equation}
In general, for $\bm{f}\neq 0$ the average velocity can always be expressed as
\begin{equation}
\langle \dot{\bm{x}} \rangle \, =  \,
\gamma_1(v_{np}) \,  v^{fr}_1 \, \hat{\bm{e}}_1 \,+\, \gamma_2(v_{np}) \,  v^{fr}_2 \, \hat{\bm{e}}_2 \, ,
\label{general}
\end{equation}
where $ \gamma_1(v_{np}) $ and $ \gamma_2(v_{np}) $ are non-linear functions of the lag whose form depend on the details of the pinning landscape $\bm{f}$.
Therefore, using \eqref{fmf_magnus2}, the mutual friction components along  $\hat{\bm{e}}_1$  and  $\hat{\bm{e}}_2$  become 

\begin{align}
\label{mf_standard2}
\begin{split}
\langle {f}^1_M \rangle 
\, & = \, 
v_{np}-\langle \dot{x}_2 \rangle 
\, = \,
\frac{1-\gamma_2+\mathcal{R}^2}{1+\mathcal{R}^2} \, {v}_{np}
\\
\langle {f}^2_M \rangle 
\, & = \, \langle \dot{x}_1 \rangle \, = \, 
\frac{\mathcal{R } \, \gamma_1}{1+\mathcal{R}^2} \, 
 {v}_{np} \, ,
 \end{split}
\end{align}
that are manifestly non-linear in $v_{np}$ due to the expected dependence of the factors $\gamma_{1,2}$ on the lag. 

%

In some pulsar glitch models \citep[e.g. ][]{haskell_probability,khomenko2018}, a non-linear mutual friction is introduced by means of formally splitting the vortex ensemble into a pinned and a free sub populations. This is equivalent to prescribing that $\gamma_{1,2}=\gamma $, where $0 \leq \gamma \leq 1 $
is interpreted as the probability of finding a free vortex moving with velocity $\bm{v}_{fr}$ in the sample \citep{jahan2006ApJ,link2014ApJ}.
Clearly, when $\bm{f}\neq 0$, a vortex can not be identified as perfectly pinned or perfectly free but, when considering a large vortex ensemble, it is natural to ask whether a mixture of pinned and free vortex lines may reproduce the correct $\langle \dot{\bm{x}} \rangle$. 

\begin{figure}
    \centering
    \includegraphics[width=0.47\textwidth]{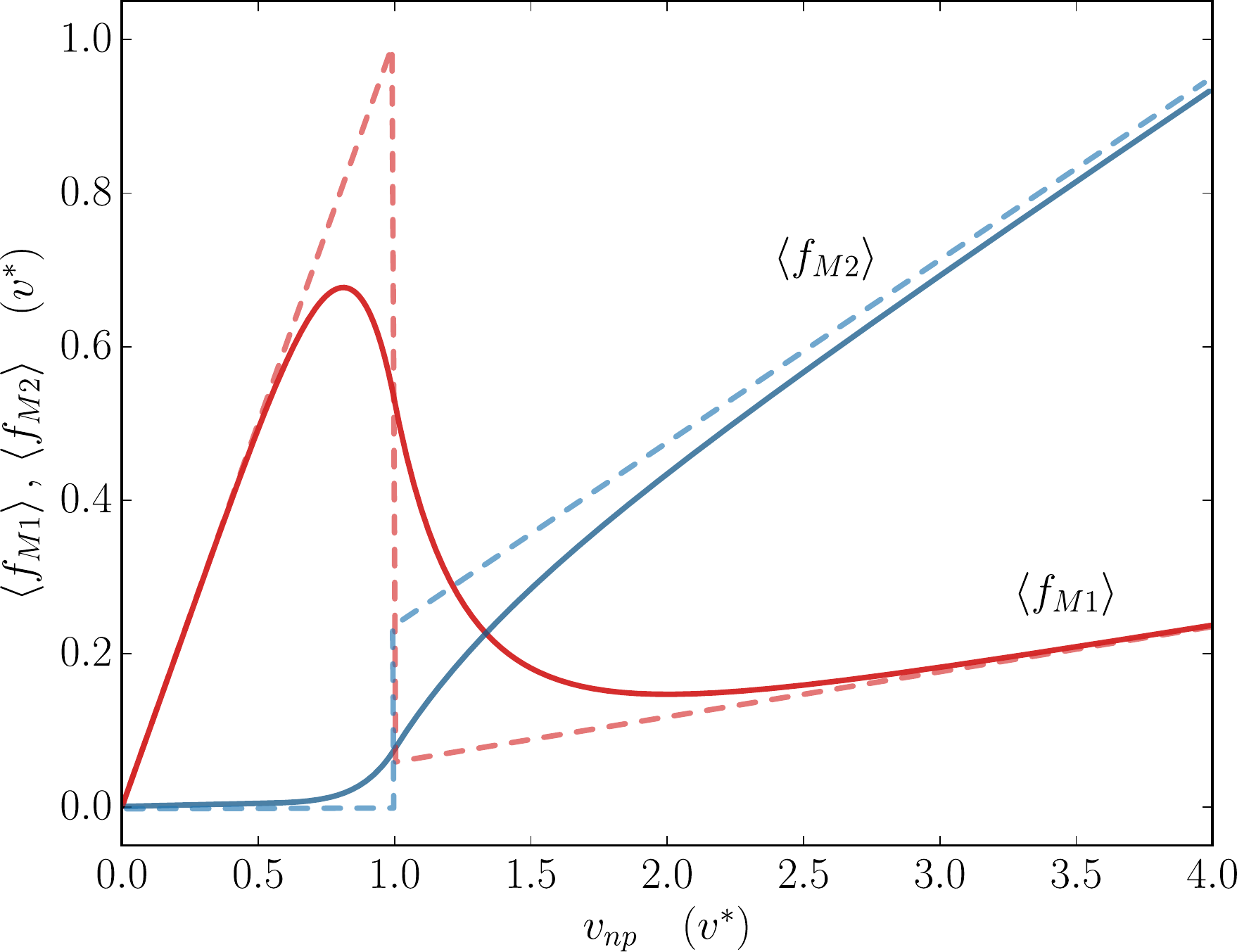}
    \caption{
    Sketch of the expected mutual friction for the minimal model in equation \eqref{overdamped}.
    The red (blue) curves refer to the component $f_{M1}$ orthogonal ($f_{M2}$ parallel) to the lag, as 
    given in \eqref{mf_standard} under the assumption that $\gamma_{1,2}=\gamma(v_{np})$.
    The dashed lines correspond to the case in which a sharp depinning transition takes place at a well defined value of the lag $v^*$, namely the unpinning probability is assumed to be a unit step function $\gamma(v_{np})=\Theta(v_{np}-v^*)$. 
    Conversely, the solid lines refer to the case of a smooth crossover between the pinned and free regimes.
    To emphasise the features in the $v_{np}\gg v^*$ regime (the linear regime), the high value $\mathcal{R}=0.25$ has been used.
    }
    \label{sketch_MF}
\end{figure}

Therefore, let us momentarily assume for simplicity that $\gamma_{1,2}=\gamma$ and briefly discuss the consequences of such an ansatz.
For the general picture of pulsar glitches based on pinning to be valid, two opposite regimes are expected. 
First, for for a large lag $ v_{np} \gg |\langle \bm{f} \rangle |$, the free-vortex regime should be recovered: in this limit $\langle \dot{\bm{x}} \rangle \approx \bm{v}_{fr}$, $\gamma \approx 1$ and the linear mutual friction \eqref{mf_standard} should be recovered. 
Conversely, for small values of $v_{np}$ the average motion of vortices is expected to be severely hindered, namely $\gamma \ll 1$. 
The simplest possibility to model this behaviour is to assume $\gamma = \Theta(v_{np} -  v^*)$, where $\Theta$ is the unit step function and $v^*$ is a critical value (to be estimated by means of microscopic arguments) of the lag below which vortices are perfectly pinned \citep{haskell2012MNRAS,antonellipizzochero2017}. 
Given this crude approximation, a sketch of the expected mutual friction  is shown in Fig \ref{sketch_MF}.

This formal splitting of the vortex ensemble into a perfectly pinned and a free population must be consistent with the fact that quantum vortices do not decay, i.e.
\begin{align}
  \partial_t n_{v} \,& = \, - \nabla \cdot(\bm{v}_{fr} \, n_{v}) \, ,
  \label{continuityNV}
\end{align}
which can be derived from the vorticity equation \eqref{hvbkn_curl2}.
Since $\gamma$ can be interpreted as the local fraction of free vortices, the effective free vortex density is $n_{fr} = \gamma \, n_v$, while  $n_{pin} = (1-\gamma)n_v$ is the complementary density of pinned vortices. Inserting this formal splitting into \eqref{continuityNV}, we must have that
\begin{align}
\begin{split}
  \partial_t n_{fr} \,& = \, - \nabla \cdot(\bm{v}_{fr} \, n_{fr}) - n_v \Gamma 
  \\
     \partial_t n_{pin}\, &= \,  n_v \Gamma \, ,    
\end{split}
  \label{continuityVort2}
\end{align}
where $\Gamma=\Gamma(v_{np},n_{fr},n_{pinn})$, that can be either negative or positive, is the net rate for the unpinning and repinning processes.

Now, at the macroscopic level the fraction $\gamma$ may locally change just  because vortices are advected with a certain average velocity that can be non-uniform. However, by restricting ourselves to a local mesoscopic domain with a conserved number of vortices $N_v = N_{pin}+N_{fr}$ and uniform lag, we can link the local values of the rate $\Gamma$ to the average motion of vortices: from  $ \dot{\gamma} = \dot{N}_{pin}/N_v =\Gamma$ we have
\begin{equation}
  \Gamma   \, = \, \dfrac{d}{dt} \, \frac{|\langle \dot{\bm{x}} \rangle|}{| \bm{v}_{fr} |} \, .
  \label{pallotto}
\end{equation}
Clearly, the physics encoded into the total rate $\Gamma$ will depend on the ingredients implemented into the description of the vortex ensemble (i.e. on the microscopic details of the dissipative processes at work, pinning and vortex-vortex interactions). 
In particular, assuming an ensemble of non-interacting vortices automatically excludes  a class of depinning processes related to vortex proximity effects \citep{melatos_Unpinning}.

Despite the idea of describing the average vortex motion by means of chemical-like reactions as in \eqref{continuityVort2} may be intriguing, in the following we will not rely on the assumption $\gamma_{1,2} =\gamma$. In fact, the components of the mutual friction will be extracted in full generality directly from \eqref{mf_standard} by calculating the average velocity components $\langle \dot{x}_1\rangle $ and $\langle \dot{x}_2\rangle$. This will allow us to perform a preliminary numerical analysis of this approximation (see Sec \ref{comparisonG1G2}).

\section{Two-dimensional pinning landscape}
 
We now consider the case in which {{the force per unit length}} $\tilde{\bm{f}}$ models the interaction of the vortex with the surrounding non-uniform medium. 
In fact, the $\tilde{\bm{f}}$ term in \eqref{overdamped} may also include also contributions other than pinning, like the the ``elastic'' contribution of the vortex array, which derives from the fact that the local velocity field around the vortex will also depend on the relative position of neighbouring vortices \citep{Haskellhop}. We shall not consider these effects in this work, so that each vortex in the ensemble is uncorrelated from the others.

Since we are considering straight vortex lines, it will be convenient to interpret {{the rescaled force $\bm{f} = \tilde{\bm{f}}/\kappa \rho_n=f_x \hat{\bm{x}} + f_y \hat{\bm{y}}$ as a fixed field (with the physical dimensions of a velocity) in the plane orthogonal to the vortex line.}} 
Moreover, the details of the inner-crust medium in which vortices are immersed are still quite uncertain \citep{chamel_review_crust}, so we will consider two different models for $\bm{f}$: one featuring quenched disorder and one describing a periodic medium (i.e. a lattice with no dislocations nor defects).


\subsection{Disordered pinscape} 
\label{disordinatoo}

It is convenient to model the effective two-dimensional pinning force field as 
$\bm{f}  \, = \, - \nabla  \Phi$, for a scalar potential of the form 
\begin{equation}
\label{potenziale2D}
 \Phi(\bm{x}) = \Phi_0 \sum_a e^{ -|\bm{x}-\bm{r}_a|^2 /  2\sigma^2  } +c \, , 
\end{equation}
where $c$ is an arbitrary constant and $\bm{r}_a$ are $N_P$ fixed random positions uniformly distributed over a large area $L^2 \gg \sigma^2 $. 
The factor $\Phi_0$ sets the strength of the pinning interaction and can be either positive or negative.
Gaussian potential wells have already been used to model the interaction of a vortex with a pinning site in the three dimensional space \citep{link2009PhRvL,wlazlowski2016}. However, in the present context equation \eqref{potenziale2D} is just a convenient prescription used to introduce some quenched disorder in the plane (see Appendix \ref{app_corr}). 
The potential \eqref{potenziale2D} is defined by three parameters, 
\begin{equation}
    \{ \,  \sigma \,  , \,  n_P= N_P/L^2 \, , \,  \Phi_0 \, \}
\end{equation}
that, leaving aside for the moment the sign of $\Phi_0$, can be cast into the equivalent set 
\begin{equation}
    \{ \,  \sigma \,  , \,    l_P= 1/\sqrt{\pi \, n_P}\, , \, v_0 = |\Phi_0|/l_p \, \} \, ,
    \label{3parametri}
\end{equation}
where $v_0$ is a velocity scale associated to the typical fluctuation of the potential and 
{{ $l_P$ is the Wigner-Seitz radius associated to the density $n_P$ in the plane.}}

\begin{figure}
    \centering
    \includegraphics[width=0.4\textwidth]{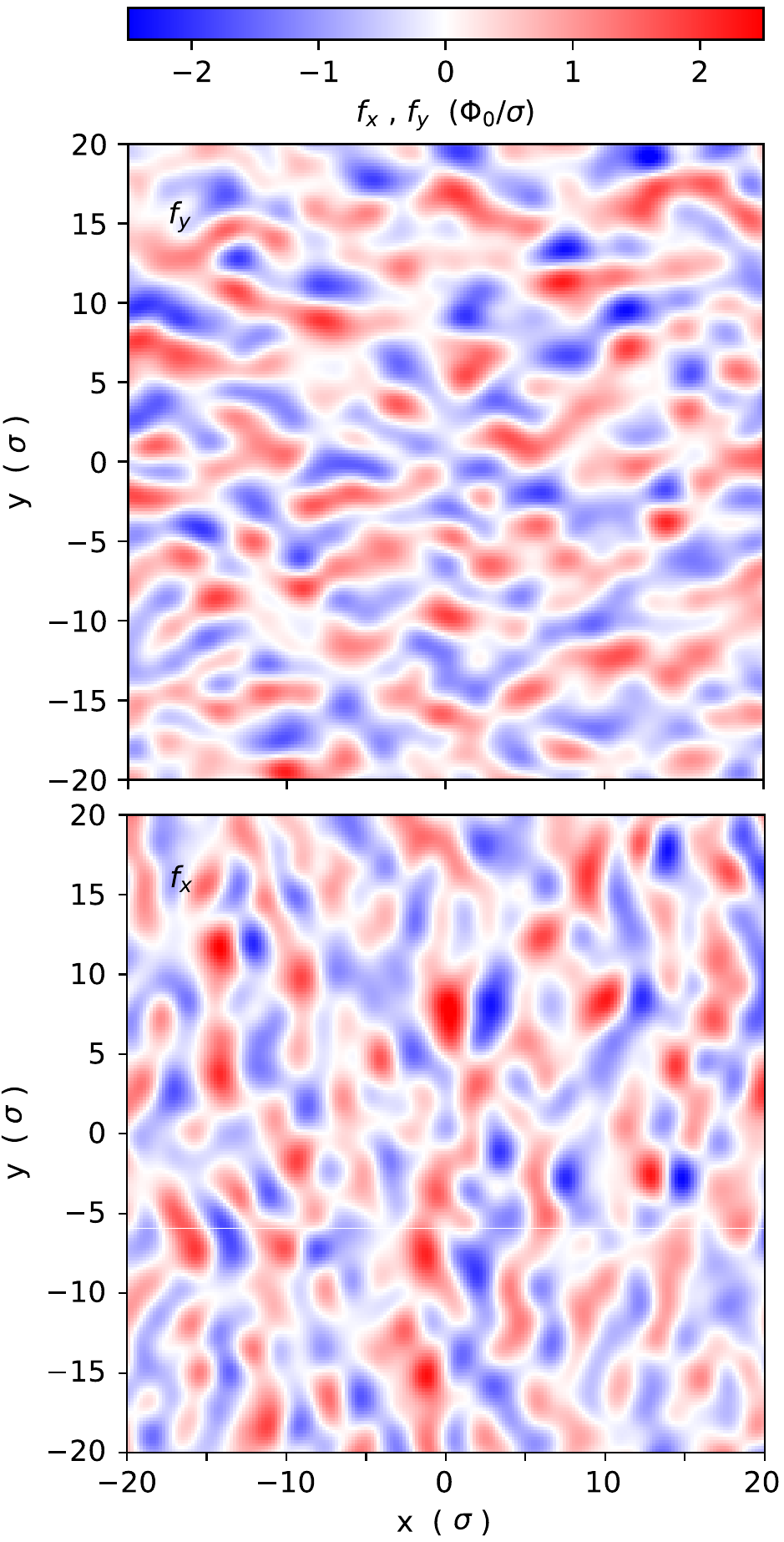}
    \caption{Example of disordered pinning landscape obtained from \eqref{forzagauss} when $\sigma=l_P$. 
    A portion of linear extension $40 \sigma$ of a larger domain is shown. The upper and lower panels refer to the components $f_y$ and $f_x$ respectively. The width of the blue and red regions measured in the direction of the component considered is set by the correlation function in \eqref{pistacchio}.
    }
    \label{fig:red_blue}
\end{figure}

{{A detailed analysis of the potential $\Phi$ is presented in Appendix B}}.
The constant $c$ in \eqref{potenziale2D} is set to the value $c =\, 2 \pi \Phi_0 n_P \sigma^2$, see \eqref{average_phiNp}, so that the average\footnote{
{{
   Here, the symbol $\langle q \rangle $ indicates an average over different realizations of the disorder, namely over the variables $\bm{r}_a$, as defined in \eqref{average_g}. In the limit of large  $N_p$ and $L$ but finite $n_P$ this averaging over the disorder is equivalent to a spatial average over the two-dimensional domain of the function $q(\bm{x})$, see Appendix \ref{app_corr}. Whether  $\langle q \rangle $ is an average over the disorder or an average over the vortex positions, i.e.  $\langle q \rangle = N_v^{-1} \sum_i q(\bm{x}_{i})$ as in \eqref{vL}, should be clear from the context. }}
} 
of the potential is zero, i.e. $\langle \Phi \rangle = 0$ in the limit of large  $N_p$ and $L$ but finite $n_P$. 
In this limit the properties of the potential are described by its two-point correlation function, {{defined in \eqref{corr}, }}
\begin{equation}
\label{gaussianC}
C(|\bm{x}-\bm{x}'|) = \langle  \Phi(\bm{x}) \Phi(\bm{x}') \rangle 
\,  = \,  
\Phi_0^2\,\sigma^2 \pi \, n_P\, e^{ -\frac{|\bm{x}-\bm{x}'|^2}{4\sigma^2}  } \, , 
\end{equation}
which tells us that the typical deviation of the potential from its average value is
\begin{equation}
\label{C0gauss}
\sqrt{C(0)}
\,  = \,  
|\Phi_0| \, \sigma \sqrt{ \pi \, n_P }
\,  = \, 
v_0 \, \sigma   \, .
\end{equation}
Given \eqref{potenziale2D}, the resulting pinning force field 
\begin{equation}
\bm{f}(\bm{x}) 
\, = \,
\frac{\Phi_0}{\sigma^2}\sum_a (\bm{x}-\bm{r}_a) e^{ -|\bm{x}-\bm{r}_a|^2 /  2\sigma^2  }
\label{forzagauss}
\end{equation}
is isotropic on large scales, i.e. $\langle \bm{f} \rangle = 0$ and its correlation function \eqref{corrFNp2} reads 
\begin{multline}
\label{pistacchio}
D_{ij}(\bm{x}-\bm{x'}) = \langle  f^i (\bm{x}) f^j(\bm{x'})\rangle 
\,  = \,  
 \Phi_0^2 \,  n_P \, \frac{\pi}{2} \, e^{ -|\bm{x}-\bm{x'}|^2 / 4\sigma^2  } 
 \\
\times \left[ \delta_{ij}-\frac{(x_i-x'_i)(x_j-x'_j)}{2 \sigma^2}  \right] \, . 
\end{multline}
A typical realization of the pinning force field is shown in Fig \ref{fig:red_blue}, for the particular case $l_P = \sigma$. 
The analytic form of $D_{ij}$ describes the geometry of the regions defined by the sign of the force components (i.e. the red and blue regions in Fig \ref{fig:red_blue}). 
{{
For example, consider the autocorrelation of the component $f_x$ by setting $i=j=x$ and $\bm{z}=\bm{x}-\bm{x}'$ in \eqref{pistacchio}. Now, $D_{xx}(\bm{z})$ is always positive if we increase the $\hat{\bm{y}}$ component of $\bm{z}$, meaning that $f_x(\bm{x}')$ and $f_x(\bm{x}'+\bm{z})$ will tend to have the same sign on a distance of a few $\sigma$. 
Conversely, $D_{xx}(\bm{z})$ becomes negative (indicating anti-correlation) by moving  a distance of $\sqrt{2} \sigma$ in the $\hat{\bm{x}}$ direction, meaning that $f_x(\bm{x}')$ and $f_x(\bm{x}'+\bm{z})$ will tend to have opposite signs for $|\bm{z}| \gtrsim \sqrt{2} \sigma$.
This explains why the red and blue regions in the lower panel of Fig \ref{fig:red_blue} extend mostly along the $\hat{\bm{y}}$ direction.
However, due to the exponential term, $D_{xx}$ drops to zero over the length scale $2 \sigma$, so that the relative sign between $f_x(\bm{x}')$ and $f_x(\bm{x}'+\bm{z})$ can assume any value for $|\bm{z}| \gg \sigma$. 
}}

\begin{figure}
    \centering
    \includegraphics[width=0.45\textwidth]{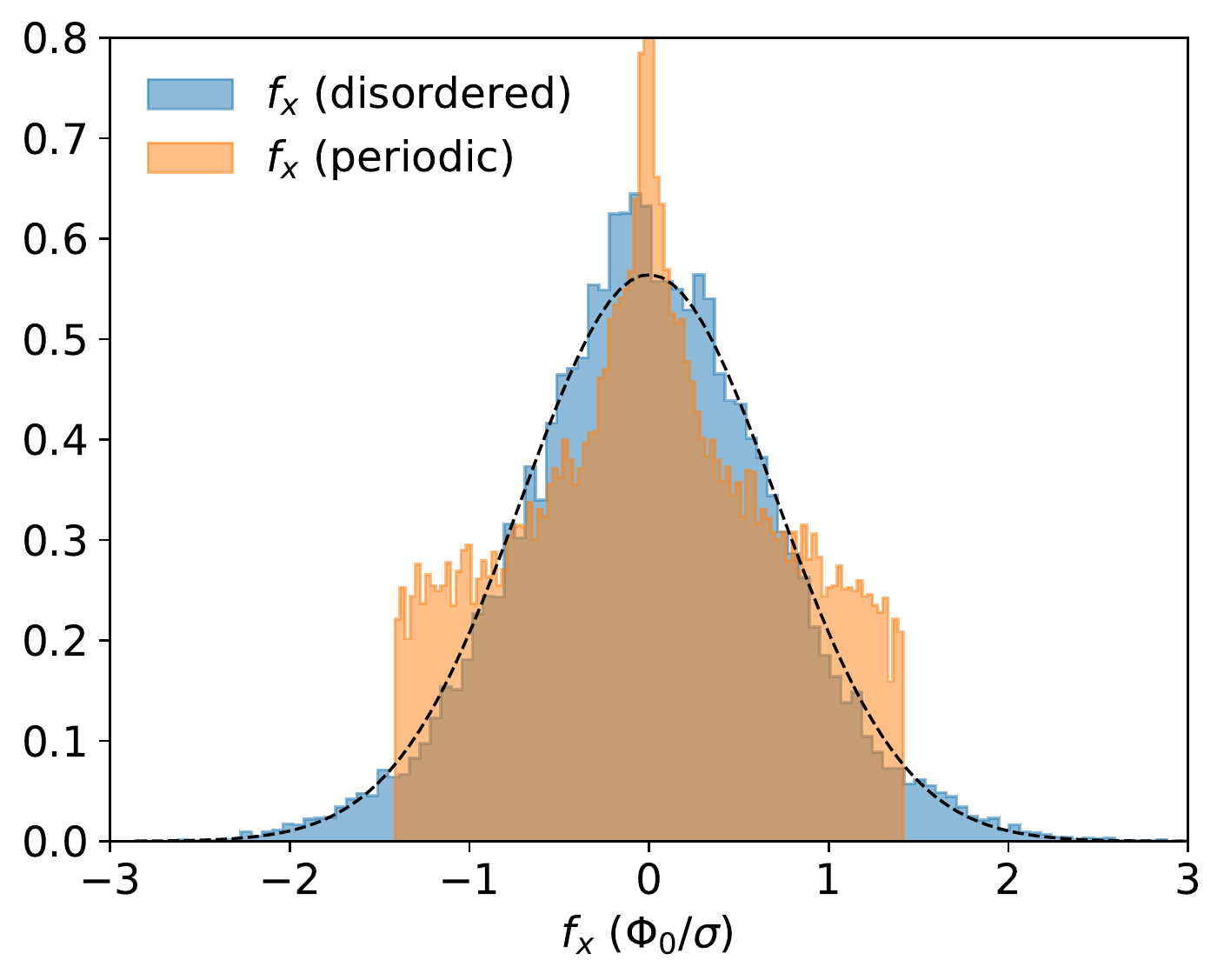}
    \caption{Distribution of the component $f_x$ of the pinning force $\bm{f}$ for the model with disorder \eqref{forzagauss}, in blue, and for the periodic model \eqref{forzanew}, in orange. 
    The particular case $\sigma=l_P$ has been considered, so that $f_x$ is given in units of $v_0=\Phi_0/\sigma$.
    The normalised histograms have been obtained by evaluating $\bm{f}$ at $10^4$ random positions uniformly distributed over the whole domain. For comparison, the expected exact distribution for the disordered pinscape (a normalized Gaussian with variance $D_{xx}(0) = v_0^2/2$)  is shown.
    }
    \label{fig:histogram}
\end{figure}

Finally, the diagonal elements $D_{ii}$ tell us that the force component $f_i$ fluctuates around zero with variance 
\begin{equation}
\label{D0gauss}
D_{ii}(0) = \langle f_i(\bm{x})  f_i (\bm{x})\rangle 
\,  = \,  
 \Phi_0^2 \,  \pi \, n_P/2 \, .
\end{equation}
This result is interesting: it is not the ratio $\Phi_0/\sigma$ that directly sets the intensity of the force, as it could seem by looking at \eqref{forzagauss}, but rather that 
\begin{equation}
|\bm{f} |   
\,  \approx \,  \sqrt{ D_{xx}(0) + D_{yy}(0) } \, = \, 
 v_0 \, .
\label{lP}
\end{equation}
Since $\bm{f}(\bm{x})$ is  the sum of many uncorrelated random variables, in the large $N_P$ limit each component $f_i$ is normally distributed around zero with a  variance $D_{ii}(0) = v_0^2 / 2 $, as shown in Fig \ref{fig:histogram}.   

\subsection{Periodic pinscape} 

Since the degree of disorder  in the inner crust is uncertain \citep{jones_amorph_99,sauls2020arXiv}, for comparison purposes we also consider a periodic potential 
\begin{equation}
\label{potenziale2Dperiodico}
 \Phi(\bm{x}) = -A \sin( \alpha x_1 )  \sin( \alpha x_2  ) \, 
\end{equation}
and the associated {{rescaled force}} $\bm{f} = -\nabla \Phi$. 
In fact, many glitch models consider an ordered potential with a regular bcc lattice geometry to represent pinning in the crust \citep[e.g. ][]{Hirasawa2001ApJ,seveso+2016}. 
A direct comparison between the two  pinscapes in \eqref{potenziale2D} and \eqref{potenziale2Dperiodico} is 
difficult because the parameters of the models have different meanings. However, it is possible to 
tune $A$ and $\alpha$ in a way that at least some average properties of the two models are the same.
{{
In this case, however, to extract these properties we can not take averages over different realizations of the disorder, as in \eqref{gaussianC}. Instead, we have to consider the usual autocorrelation for periodic signals,
}}
\begin{align}
\label{corr2Dperiodico}
\begin{split}
 C(\bm{z})  &= \frac{\alpha^2}{(2\pi)^2} \int d^2r \, \Phi(\bm{z}+\bm{r})\Phi(\bm{r}) 
 \\
   &= \frac{A^2 }{4}   \cos( \alpha z_1 )  \cos( \alpha z_2   ) \, ,
\end{split}
\end{align}
where the integral is performed over a square of side $2 \pi/\alpha$.
The autocorrelation of the {{rescaled force $\bm{f}$}} is obtained as 
\begin{align}
\label{Dperiodico}
\begin{split}
 D_{ij}(\bm{z})  = &-\frac{\partial^2}{\partial z_i\partial z_j} C(\bm{z})
 \\
   = &\, \frac{A^2 \alpha^2}{4}   [ \,  \delta_{ij}  \cos( \alpha z_1 )  \cos( \alpha z_2   ) + 
   \\
   & \, \, \,   +( \delta_{ij}-1)  \sin( \alpha z_1 )  \sin( \alpha z_2   ) \, ] \, .
\end{split}
\end{align}
Similarly to \eqref{C0gauss} and \eqref{D0gauss}, we have that the potential and the force fluctuate with a variance of 
\begin{align}
\label{varPeriod}
  C(0)  = \frac{A^2}{4} \qquad 
  D_{xx}(0)  =  D_{yy}(0)  = \frac{A^2 \alpha^2}{4} \, .
\end{align}
Therefore, to compare the effect of the disordered and periodic pinning landscapes on the vortex motion, we set $A$ and $\alpha$ in such a way that  $C(0)$ and $D_{ii}(0) $ are equal to the ones given in \eqref{C0gauss} and \eqref{D0gauss}, namely we impose  
\begin{align}
\label{varPeriod}
  \alpha  = \frac{1}{\sqrt{2} \, \sigma}  \qquad A  = \frac{2 \, \sigma  \Phi_0 }{l_P}\, .
\end{align}
In this way the resulting pining force is parametrized as 
\begin{equation}
\label{forzanew}
 f_i(\bm{x}) =  \sqrt{2}  \, v_0 \,  \cos\left( \frac{x_i}{\sqrt{2} \,  \sigma } \right)  
 \sin \left(  \frac{x_k}{\sqrt{2} \, \sigma }  \right)  \,  \qquad k\neq i \, .
\end{equation}
Figure \ref{fig:histogram} shows a comparison between the distributions of the values $f_i$ arising from the disordered model \eqref{forzagauss} and the periodic model \eqref{forzanew} tuned according to \eqref{varPeriod}. Now, the $f_i$ is not the sum of many independent random variables, so its distribution is not Gaussian, as in the disordered case. However, its average value is still $\langle f_i \rangle =0$ and the variance coincides with the one of the Gaussian.
Moreover, since  an hypothetical critical lag for unpinning cannot exceed the maximum of $f_i$, namely $ \sqrt{2}  v_0 $, we automatically know that for this periodic pinning landscape the critical lag for unpinning $v^*$ will be $v^* < \sqrt{2} \Phi_0/l_P $. This can be clearly seen in Fig \ref{fig:histogram}, where the distribution of  $f_i$ drops to zero for  $|f_i|>1.41 \Phi_0/l_P$.

\begin{figure}
    \centering
    \includegraphics[width=0.46\textwidth]{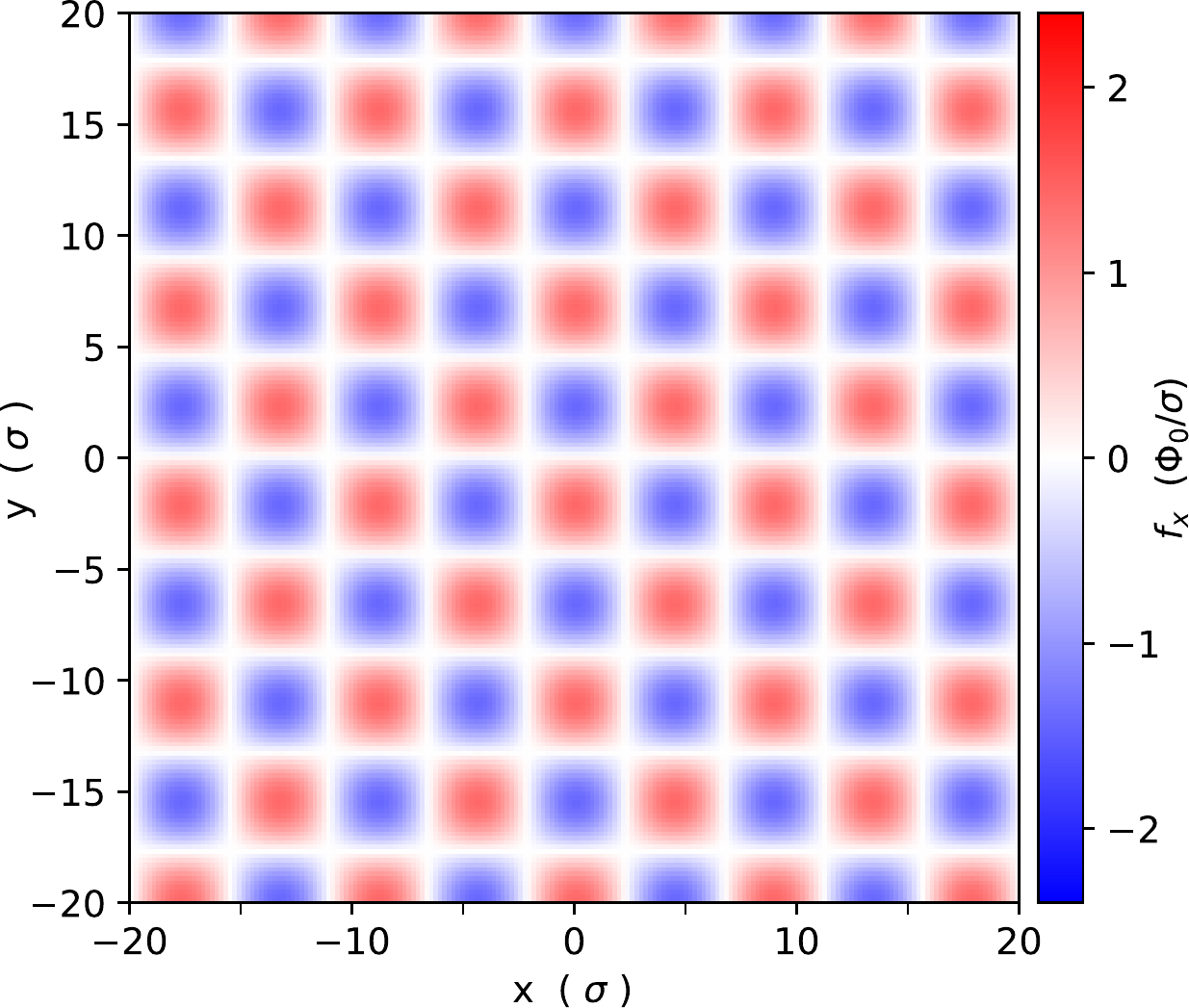}
    \caption{The component  $f_x$ of the periodic pinning landscape obtained 
    from \eqref{forzanew} when $\sigma=l_P$. 
    For better comparison with Fig \ref{fig:red_blue}, the same colour scheme is used (despite the fact that $|f_x|<1.41 v_0$ in this case, see Fig \ref{fig:histogram}) and a domain of identical extension is shown. 
    The linear extension of the blue and red regions is $\pi\sqrt{2}\, \sigma \approx 4.4 \, \sigma$.
    }
    \label{fig:red_blue_P}
\end{figure}

{{Finally, to better compare the periodic pinscape geometry with the disordered one, the  component $f_x$ of the rescaled pinning force defined in \eqref{forzanew} is plotted in Fig \ref{fig:red_blue_P}.
}}

\subsection{Typical parameters in a neutron star crust}
\label{sec_parameters}

Before moving to the numerical analysis of the system it is worth discussing the typical 
values of the three phenomenological parameters in \eqref{3parametri}. In the following, we will explicitly refer to the disordered pinscape \eqref{forzagauss}. However, thanks to the tuning of the parameters in \eqref{varPeriod}, the discussion is also valid for the periodic pinscape \eqref{forzanew}.

First, the {{rescaled pinning force $\bm{f}=\tilde{\bm{f}}/\kappa\rho_n$}} depends on two length scales, $\sigma$ and $l_P$. 
In the case in which $l_P \gg \sigma$, the potential $\Phi$ in \eqref{potenziale2D} consists of many scattering centers, surrounded by regions in which the force is almost zero. 
In this limit, the study of a vortex scattering off a single potential well \citep{sedrakianRepinning} can be used to investigate whether a vortex that unpins in a realistic NS setting would re-pin before encountering another vortex \citep{Haskellhop}.

However, this is not the situation we expect in neutron star interiors, where the vortex core radius is comparable to the expected Wigner-Seitz radius of nuclei and the vortex may remain straight over many crystal domains \citep{link2009PhRvL}: both these facts contribute to smear and renormalize the effective two-dimensional pinning potential of an extended vortex segment \citep{seveso+2016}. 
Moreover, it is also possible that the potential $\Phi$ would arise from the interaction with a kind of pasta phase rather than with nuclei organized in crystal domains.

To date this effective potential is very uncertain and it is not obvious how to express its properties starting from the knowledge of the  energetics \citep{donati2004} and the dynamics \citep{bulgac2013prl,wlazlowski2016} arising from the interaction of a vortex with a single nucleus.
However, it is still possible to give at least a rough estimate of $\sigma$, $l_P$ and $v_0$.

From the practical point of view, in order to create an effective potential in which the single Gaussian wells are not clearly identifiable as separate scattering centers, we will set $\sigma = l_P$ in the rest of this work. 
Despite the single Gaussians may have a definite attractive or repulsive character depending on the sign of $\Phi_0$, the full potential is neither attractive nor repulsive. 
In fact, the inversion $\Phi_0 \rightarrow -\Phi_0$ does not change dramatically the overall shape of the potential  when the average distance $l_P$ between the Gaussian centers is comparable to $\sigma$ \citep{link2009PhRvL}.

\begin{figure}
    \centering
    \includegraphics[width=0.45\textwidth]{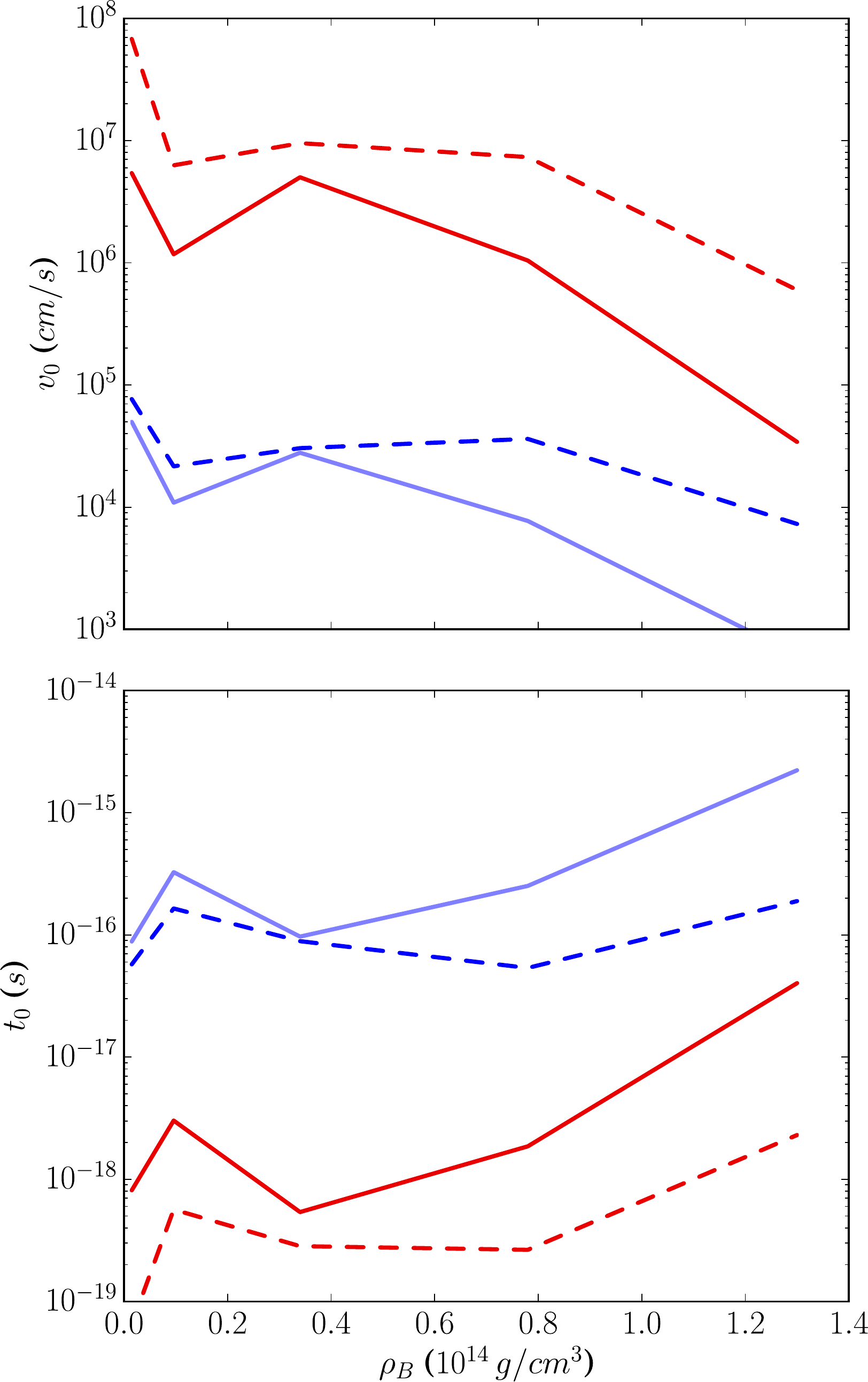}
    \caption{The phenomenological parameters $v_0$ and $t_0$ in the inner crust, according to the estimates given in \eqref{ws_estimate}, red curves, and \eqref{seveso_estimate}, blue curves.
    The solid lines refer to the case $\beta = 1 $ in \citet{donati2006}, see also \citet{seveso+2016}, namely no reduction of pairing from the polarization of the strongly correlated neutrons.  
    For comparison, also the case $\beta = 3 $ has been considered (dashed curves). To bracket all the possible values considered in the literature, the case $N=5000$ is used for the blue curves (see Tab 3 in \citealt{seveso+2016}).
    }  \label{fig:units}
\end{figure}

Setting  $l_P=\sigma$  in \eqref{forzagauss} and \eqref{forzanew}, the pinning field depends on two parameters only: the velocity  $v_0 = |\Phi_0|/\sigma$ and the  length scale  $\sigma$. 
Hence, a natural time unit for the dynamics of a vortex in such a potential is $t_0 = \sigma^2 /|\Phi_0|$.

To estimate $\sigma$, $v_0$ and $t_0$ we take the total energy difference  $\Delta E_\beta$ between two reference configurations of a vortex segment of length $2R_{WS}$, where $R_{WS}$ the Wigner-Seitz radius in the inner crust\footnote{
{{In principle, we distinguish between $l_P$, which is a parameter of the effective two-dimensional pinning potential, and $R_{WS}$, which is a physical property of the three-dimensional solid in the crust. Because of the choices made to fix the phenomenological parameters of the potential, their values coincide in the present analysis.}}
}: one configuration in which the vortex is superimposed to a nucleus and one in which it passes through the boundary between two adjacent Wigner-Seitz cells \citep{donati2004}. 
The energy difference  $\Delta E_\beta$  depends on an uncertain  positive parameter $\beta=1,3$ related to the reduction of pairing expected from the polarization of the strongly correlated neutron medium \citep{donati2006}, that can have a significant impact on the calculated values of $\Delta E_\beta$. 

The typical force $F_\beta$ experienced by a vortex segment of length $\sim 2R_{WS}$ is therefore $F_\beta \approx \Delta E_\beta/R_{WS}$. Therefore, setting $\sigma \approx R_{WS}$ gives
\begin{equation}
\label{ws_estimate}
    v_0 \approx  \frac{\Delta E_\beta}{2 \kappa \rho_n R_{WS}^2} 
    \qquad 
    t_0 = \frac{\sigma}{v_0} \approx \frac{2 \kappa \rho_n R_{WS}^3}{\Delta E_\beta} \,.
\end{equation}
These two quantities are shown in Fig \ref{fig:units} as functions of the baryon density $\rho_B$ in the inner crust. 
The free neutrons density $\rho_n$ and $R_{WS}$ have been taken from \cite{negele1973} and $\Delta E_\beta$ from \cite{donati2006}. 
Considering that $\rho_n\sim 10^{14}$g/cm$^3$, $R_{WS}\sim 20\div 40\,$fm and $\Delta E_{\beta}\sim 1\,$MeV in most of the inner crust, we have that $v_0 \sim 10^7 \, $cm/s, in broad agreement with the estimate of the microscopic vortex velocity scale derived by considering the Bernoulli force exerted on a vortex by a nucleus \citep{gerci_alpar_2016}. 

An alternative to this Wigner-Seitz approach, where the effective rigidity of the vortex is taken into account, has been proposed in \cite{seveso+2016}. In this case the mesoscopic pinning force per unit length has been calculated by considering the energetics of a straight vortex segment immersed in a crystal domain of length $ N R_{WS} \sim 10^3 R_{WS}$. Since the interactions with the single nuclei tend to cancel out, the resulting pinning force per unit length $f_{N \beta}$ turns out to be $f_{N \beta} \sim 10^{-(3\div 1)} F_\beta/R_{WS}$, the exact values depending on the values of  $N$ and $\beta$ considered.
In this case, 
\begin{equation}
\label{seveso_estimate}
    v_0 \approx \frac{f_{N \beta}}{ \kappa \rho_n } 
    \qquad 
    t_0 \approx \frac{\kappa \rho_n R_{WS}}{f_{N \beta}} \, ,
\end{equation}
which is shown in Fig \eqref{fig:units} for the cases $\beta=1,3$ and $N=5000$ \citep[see Tab 3 in][, where $N$ is referred to as $L$]{seveso+2016}.
Despite the large uncertainties,  in the whole inner crust the typical timescale $t_0$ needed for a vortex to move a distance $\sim R_{WS}$ because of velocity fluctuations induced by the pinscape is  $t_0 \sim 10^{-16}\,$s or smaller. Such a fluctuating fast motion of zero average velocity (if the lag is zero) happens on a timescale that is separated by more than ten orders of magnitude with respect to the modulations of the lag $v_{np}$ during the spin up phase in a glitch, which is expected to occur on the timescale of a second \citep{ashton2019NatAs,montoli2020A&A}. For this reason it makes sense to study the dynamics of the vortex ensemble for a fixed value of the external lag $v_{np}$, or for very slow modulations of the lag.

%
%

\section{Numerical analysis}
\label{sec:results}

Equation \eqref{normal} is solved numerically for $N_v$ non-interacting vortices, distributed over a two-dimensional domain of size $L \times L$ with periodic boundary conditions. 
In this way the ensemble explores different parts of the pinscape and, in the limit $L \gg l_P$, there is no need to average over different realizations of the disordered pinning potential: sub-domains of the pinscape that are more than a few times $\sim 2 \sigma$ apart  will tend to be uncorrelated and can be considered as belonging to different realizations of the disorder.

To set the computational domain we fix $N_P=4\times 10^4$, which gives $L \approx 355 \, l_P$. For the periodic pinning landscape in \eqref{forzanew}, a smaller domain of linear size $L=2\sqrt{2} \pi \sigma$ encompassing a single period is sufficient.

It is worth mentioning that, given the above setting, the typical distance between two vortices is $l\sim L/\sqrt{N_v} \sim 30\div 100 \, l_P$, while in real neutron star (if we tentatively identify $N_P$ with the number of pinning centers in our domain), a larger  $l\sim L/\sqrt{N_v} \sim 10^7 l_P$ is expected. 
In fact, the typical distance between ions in the crust is $\sim 10^{-10}\,$cm, while vortices are   roughly $10^{-3}\,$cm far apart for a $10\,$Hz pulsar \citep{Haskellhop}. 
However,  since we work with periodic boundary conditions, the computational domain is infinite and we can think the non-interacting point vortices as being very distant from one another, in different computational $L\times L$ cells that tessellate the plane.  
While this is not a problem for the periodic potential, it represents a possible point of concern when  the random pinning potential is used, that is now periodic with period $L$. In fact, the tessellation introduces a preferred orientation in an object (the disordered pinning potential) that should not have any preferred direction.

A solution to this problem (alternative to the implementation of unpractical very large domains) stems from the fact that, although little is known about the defect structure of the crust, one does not in general expect the pinning landscape orientation to have anything to do with the local lag direction $\hat{\bm{e}}_2$, even for a perfectly regular crystal \citep{seveso+2016}. 
Hence, we impose that each vortex (labelled by $i=1...N_v$) experiences the same strength of the background lag $v_{np}$, but with a different (constant in time) orientation $\hat{\bm{e}}_2^i$, i.e.
\begin{equation}
    \bm{v}^i_{np }\,  = \, v_{np} \, \hat{\bm{e}}_2^i = v_{np} (\cos \beta_i \, \hat{\bm{x}} + \sin \beta_i \, \hat{\bm{y}} ) \, ,
\end{equation}
where the random $\beta_i$ are uniformly drawn in $[0,2\pi)$. 
In this way it is very unlikely that the vortices will tend to follow a preferred path in the computational domain (i.e. a particularly favourable valley in the potential aligned almost along ${\bm{v}}_{fr}$), because they will be dragged by the lag in different directions, some along the favourable path at the bottom of a valley, some against the walls of the valley itself.

Preliminary numerical tests have been performed to asses the robustness of our results when $L$ is varied: qualitatively the results are always the same for $L \geq 10 \, l_P$ and we found no quantitative differences for $L \geq 100 \, l_P$. This means that for  $L \geq 100 \,  l_P$, different realizations of the disorder do not give rise to appreciable differences in $\langle \dot{\bm{x}}(t)\rangle $.


Finally, the following details are common to all the simulations performed. 
The pinscape is initialized by computing and storing the values of $\bm{f}$ in a regular two-dimensional grid 
with $\sim 10 L/l_p$ points in each dimension ($\sim 10^3 L/l_p$ if the periodic potential is used). 
At each time step, bilinear interpolation is used to compute the pinning force at vortex positions. The contributions to the pinscape from the Gaussians close to the borders are summed in a way that in the end the grid matches the periodic boundary condition requirement. 

At $t=0$, the random initial positions $\bm{x}_i(0)$ are uniformly drawn in the  $L\times L$ domain for $i=1,..., N_v$. 
For $t>0$, the trajectories $\bm{x}_i(t)$ are evolved with the Adams-Bashforth linear multistep method of the fifth-order with a constant time step $\delta t = 10^{-3} \sigma^2 / \Phi_0$. All the results are unchanged if the time step is increased to $\delta t = 10^{-2} \sigma^2 / \Phi_0$.
During the evolution, we keep track of the instantaneous velocity $\dot{\bm{x}}_i(t)$, so that 
the components $ \langle \dot{x}_j(t)$ of $\langle \dot{\bm{x}} \rangle$ in the physical basis \eqref{basis} are extracted as
\begin{equation}
    \langle \dot{x}_j(t) \rangle \, = \,  N_v^{-1} \sum_i \, \hat{\bm{e}}_j^i \cdot \dot{\bm{x}}_i(t)
    \qquad \text{for } j=1,2
    \label{basis12}
\end{equation}
at each time step.

{{Finally, we recall that, since we will set $l_P=\sigma$ in the numerical simulations (as discussed in Sec \ref{sec_parameters}), then all the velocities and the rescaled forces, like the Magnus lift $\bm{f}_M$ and the pinning force $\bm{f}$, will be given in units of~$v_0=\Phi_0/\sigma$, see Fig \ref{fig:units}. }}

\subsection{Relaxation towards the pinned state }

A first natural question is how fast a vortex randomly placed in the potential will pin. 
This kind of preliminary test is interesting from the theoretical point of view but it is better to keep in mind that in a real system this way of choosing the initial condition is highly unrealistic, as vortices happen to be at a specific position because of their past history. Therefore, this test is useful to see how fast a strongly out-of equilibrium initial configuration (where the initial positions are completely uncorrelated with respect to the pinning potential) relaxes. 

\begin{figure}
    \centering
    \includegraphics[width=0.45\textwidth]{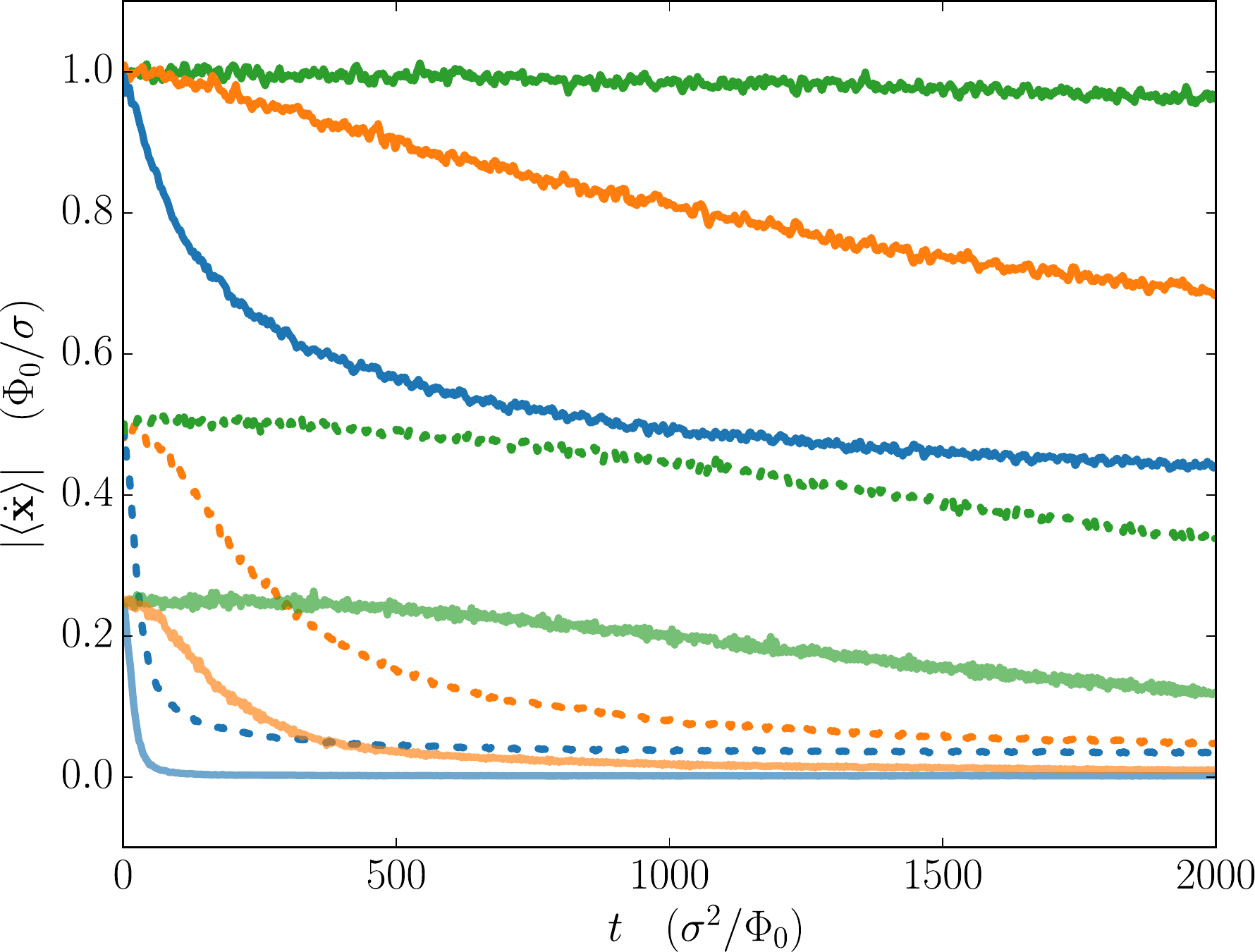}
    \caption{Temporal evolution of the modulus of the average velocity of $N_v=10^4$ vortices in the periodic pinning force field \eqref{forzanew} for three different background lags and three different values of the drag-to-lift ratio: $\mathcal{R}=0.1$ (blue), $\mathcal{R}=10^{-2}$ (orange), $\mathcal{R}=10^{-3}$ (green). 
    As discussed in the text,  $\sigma = l_P$ has been used, so that  velocities are in units of $v_0=\Phi_0/\sigma$ and time in units of $t_0 = \sigma^2/\Phi_0$.
    The curves referring to certain value of the lag are easy to identify, as they all have the same height $|\langle \dot{\mathbf{x}} \rangle  |=\cos \theta_d \,  v_{np} \approx v_{np} $ at $t=0$.  The constant lags used are: ${v}_{np}=v_0$ (the solid dark lines), ${v}_{np}=0.5 v_0$ (dotted lines) and ${v}_{np}=0.25 v_0$ (solid light lines).
    }
    \label{fig:rep_per}
\end{figure}

\begin{figure*}
    \centering
    \includegraphics[width=0.99\textwidth]{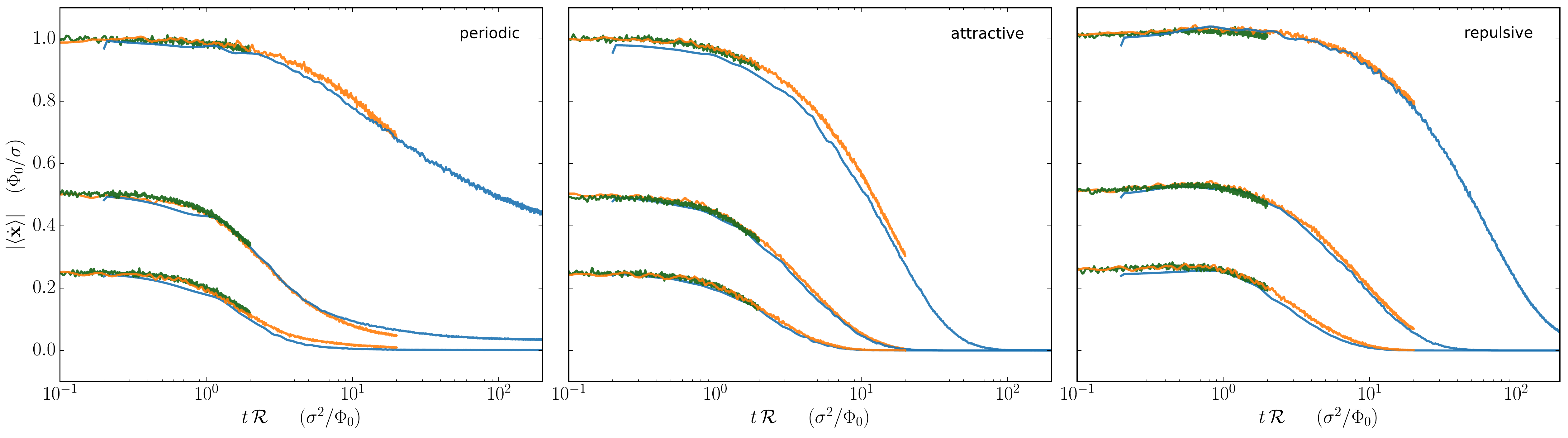}
    \caption{
    Relaxation towards a pinned state for the periodic (left), disordered ``attractive'' ($\Phi_0<0$, center)  and disordered ``repulsive'' ($\Phi_0>0$, right) potentials. 
    The modulus $|\langle \dot{\bm{x}} \rangle|$ of the vortex average velocity is plotted against the rescaled time $t\mathcal{R}$ for three different values of the drag-to-lift ratio: $\mathcal{R}=0.1$ (blue), $\mathcal{R}=10^{-2}$ (orange), $\mathcal{R}=10^{-3}$ (dark green).
    The lags used are $\bm{v}_{np}=1, 0.5, 0.25\, v_0$. The curves in the left panel coincide with the ones in Fig \ref{fig:rep_per}. Velocities  are in units of $v_0=\Phi_0/\sigma$ and time in units of $t_0 = \sigma^2/\Phi_0$.
    }
    \label{fig:rep1}
\end{figure*}

An example of this kind of relaxation for the periodic potential is shown in Fig \ref{fig:rep_per}, where $| \langle \dot{\bm{x}}(t)\rangle |$ is plotted for different values of the drag $\mathcal{R}$ and of the lag $v_{np}$.
The figure shows that, as vortices tend to pin, the velocity calculated on the initial vortex configuration decreases from the initial value $| \langle \dot{\bm{x}}(0)\rangle | = |\bm{v}_{fr}|$ on different timescales, depending on the lag and drag used.
This can be understood by recalling that at $t=0$ the $N_v$ initial positions ${\bm{x}}_i(0)$ are sampled randomly, and so the forces $\bm{f}({\bm{x}}_i(0))$. Therefore, according to Fig \eqref{fig:histogram}, we must have that $\langle \bm{f}(0) \rangle  \approx 0$, with a standard deviation of $~|\Phi_0|/(l_P \sqrt{N_v})$.
Hence, fluctuations of the order of $\sim 0.01 |\Phi_0|/l_P$ around the value $\langle \dot{\bm{x}} \rangle = \bm{v}_{fr}$ at $t\approx 0$ are a consequence of the finite number of vortices used ($N_v = 10^4$ here).


It is interesting to seek whether the  curves in Fig \eqref{fig:rep_per} obey some scaling property. 
A possibility would be to check if a rescaling of time  $t \rightarrow t\mathcal{R}$ can account for the different slopes of the curves for a fixed value of $v_{np}$. 
Such a scaling is indeed expected since, for $v_{np}=0$, the drag parameter sets the angle $\theta_d$ that the vortex trajectory makes with the level sets $\Phi=const$. 
To see this, consider $v_{np}=0$ and the small angle expansion $R_\alpha \approx 1 +\alpha R_{\pi/2}$, so that 
\begin{equation}
\label{nolag}
\dot{\bm{x}} \, = \, -\cos \theta_d \, 
  R_{\frac{\pi}{2}-\theta_d}    \nabla \Phi
  \, \approx \,  \, -R_{\frac{\pi}{2}} \nabla \Phi - \mathcal{R} \nabla \Phi \, 
 \, .
\end{equation}
The term $R_{\pi/2} \nabla \Phi $ is parallel to the $\Phi=const$ lines and, when averaged over many vortices, gives a negligible contribution to the average velocity since the level sets of $\Phi_0$ are almost always closed loops. Hence, we are left with 
\begin{equation}
\label{nolag2}
\langle \dot{\bm{x}} \rangle \, \approx  \, -\mathcal{R} \langle \nabla \Phi \rangle =  \mathcal{R} \langle \bm{f} \rangle\, 
 \, .
\end{equation}
so that $\mathcal{R}$ can be adsorbed into a rescaling of time.

This is verified numerically in Fig \ref{fig:rep1}: the curves in the left panel coincide with the ones in Fig \ref{fig:rep_per}, while the other two panels show the analogous cases for a disordered  potential made of ``attractive'' Gaussian wells  (i.e. $\Phi_0<0$) and a ``repulsive'' one ($\Phi_0>0$).
The simulations show that the expected scaling behaviour is recovered with a good approximation for all the lags and for all the different potentials tested, confirming that the scaling is a universal property of this kind of systems (i.e. independent on the potential used).

\subsection{Response to slow lag variations: hysteresis }

We now want to understand if there exist a well-defined depinning threshold corresponding to a critical value $v_{np}=v^*$ above which a sample of pinned vortices starts to move with a non-zero average velocity.

To investigate this possibility we start with a pinned vortex configuration and slowly increase the lag, thus mimicking the effect of gradual spin-down in a pulsar. 
Hence, the simulation is pre-initialized with vortex lines in random positions and $\bm{v}_{np}=0$. Each vortex is then evolved till its velocity is zero, so that it relaxes to the closest pinned position. 
Since the final pinned position does not depend on the value of  $\mathcal{R}$, we can impose $\mathcal{R}=1$ during this preliminary phase to speed-up this  repinning process. 
After this preliminary procedure at $v_{np } =0$ is completed, the drag  $\mathcal{R}$ is set to the actual value we want to consider and the real evolution starts: we set $t=0$ and the modulus of the input lag is slowly modulated for $t>0$ as 
\begin{equation}
    v_{np}(t) = v_{np}^{\rm{max}} \sin{(2\pi t/T)} \, ,
    \label{modulation}
\end{equation}
where $T$ is the period of the process. 
As seen in the previous subsection, due to the universal scaling property shown in Fig \ref{fig:rep1} the minimal requirement for \eqref{modulation} to be a slow modulation  is that $T \gg t_0 / \mathcal{R} $. 
This condition is certainly met in a real pulsar, given the values of $t_0$ in Fig \eqref{fig:units}.

\begin{figure}
    \centering
    \includegraphics[width=0.47\textwidth]{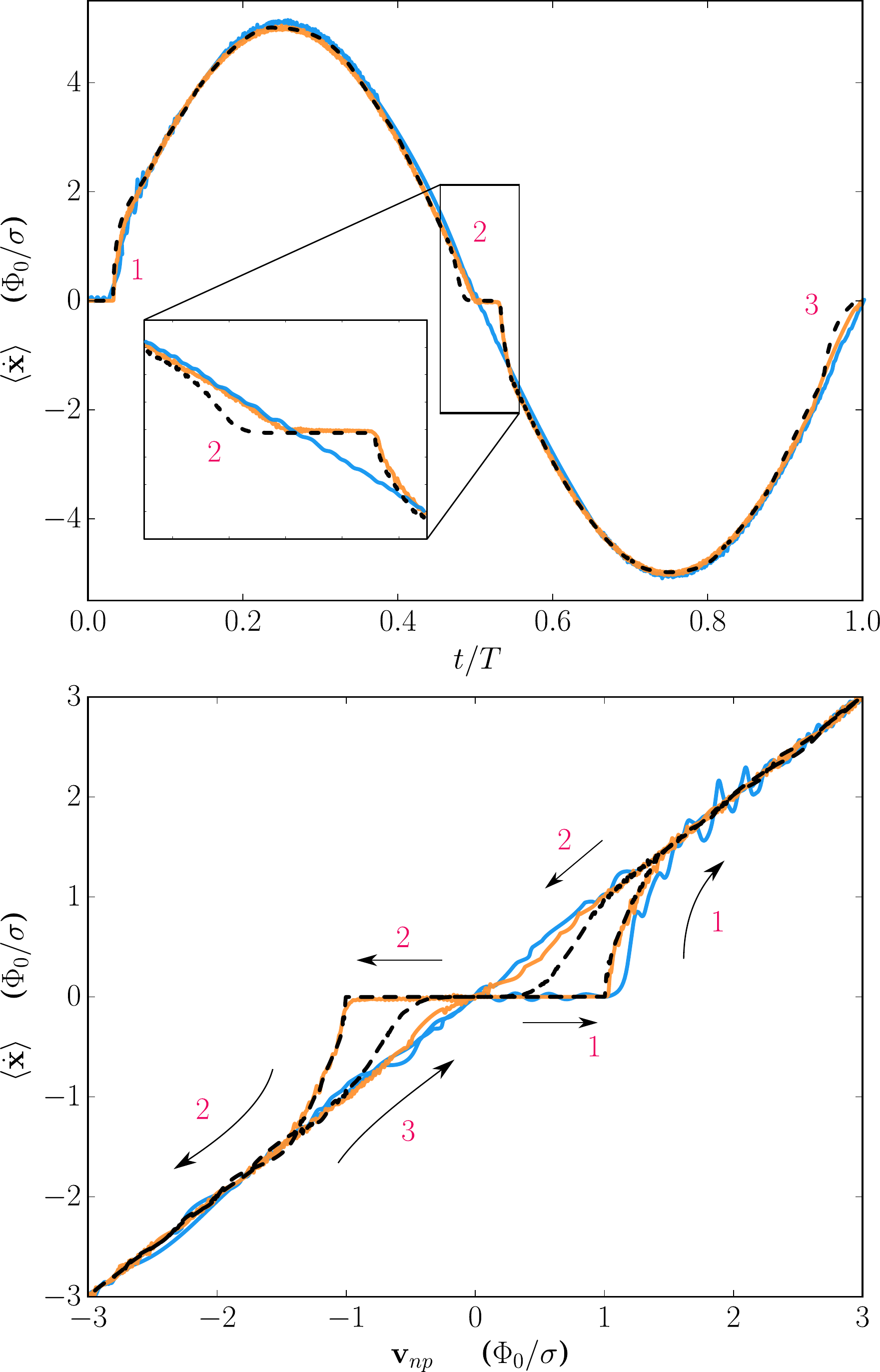}
    \caption{
    Upper panel - Evolution of the average velocity component $\langle \dot{x}_2\rangle$ of  $N_v=10^3$ vortices for $\mathcal{R} = 10^{-2}$. The periodic potential has been used.
    Both $\langle \dot{x}_2\rangle$ and $v_{np}$ are in units of $v_0=\Phi_0/\sigma$.
    The lag is modulated according to \eqref{modulation} with $v_{np}^{\rm{max}} = 5 v_0$ and $T = 10^3 t_0$ (blue curve), $T = 10^4 t_0$ (orange curve), $T = 10^5 t_0$ (black dotted curve).  
    Lower panel - The same evolution is plotted against the instantaneous lag speed $v_{np}(t)$. 
    The arrows indicate the orientation of the hysteresis cycle and the red numbers refer to the three different phases of the evolution shown in the upper panel. For all the values $T$ tested, a sharp depinning transition  occurs at $v^* =  v_0$ during phase-1. 
    To make the hysteresis loop more evident, the curves have been filtered to remove most of their intrinsic noise. The drawback of this are the artificial oscillations that can be seen in the blue curve (the one corresponding to the shortest $T$, that is the most noisy) during the late phase-1 in the lower panel.
    }
    \label{fig:sinY}
\end{figure}

To set a value $T$ that is long enough (but not impractically long) in our simulations we performed some  preliminary test, an example of which is shown in Fig \ref{fig:sinY}. It can be seen how the velocity component parallel to the lag, i.e. $\langle \dot{x}_2(t) \rangle$ as defined in \eqref{basis12}, follows the sinusoidal modulation in \eqref{modulation} for different values 
of $T$ and $\mathcal{R}=10^{-2}$.

In the first phase of the evolution shown in Fig \ref{fig:sinY} (phase-1), the vortex velocity remains zero for all the values of $T$ tested until a certain critical lag $v^*$ is reached. Then, vortices unpin and  $\langle \dot{x}_2(t) \rangle$ quickly adjusts to the curve $v^{fr}_2 \approx v_{np}(t)$. 
This means that, once the lag $v_{np}(t)$ overcomes the critical value $v^*$, the ensemble enters into a phase in which  vortices almost move (on average) with the free  velocity $\bm{v}_{fr}$ defined in \eqref{basis2}.
This lasts till  $v_{np}(t)$ drops again below $v*$: for $T=10^5t_0$ the vortices tend to repin, but in a smoother way than the unpinning of phase-1. When they unpin again, for $v_{np}(t) \approx -v^*$, the same abrupt behaviour of phase-1 is recovered. Therefore, it seems that the unpinning and the repinning behave differently, so that the periodic modulation of the lag gives rise to an hysteresis loop. 
This signals that the system retains some memory of the past, but a limited one because it disappears as the output is varied more slowly. Hence, we can identity this behaviour as a kind of rate-dependent hysteresis, which is a quite common property of dissipative driven systems. 
In fact, we find that the area of such an hysteresis loop decreases with increasing $T$, namely the depinning and repinning paths are more similar when $v_{np}(t)$ is varied slowly (compare e.g. the phase-1 and phase-2 in Fig \ref{fig:sinY}).

The presence of rate-dependent hysteresis  has also been observed in the motion of quantized flux-tubes in superconducting systems \citep{fily2010PhRvB}. 
The reason behind this hysteretic response can be understood by examining Fig  \eqref{fig:rep_per}, where it can be seen that the relaxation timescale of an out-of equilibrium ensemble grows with the lag $v_{np}$. Loosely speaking,  vortices coming from a situation in which the lag is high relax more slowly (i.e. have a longer memory of their slightly out-of equilibrium state) than vortices coming from a situation in which the lag is small but slowly increasing (which is the situation during the depinning process). 

It is noteworthy that during phase-1 (and only during phase-1) all the cases behave in the same way. This is due to the initialization procedure needed to prepare the initial pinned configuration and is in  accordance with the intuitive explanation of the hysteretic response. 

Despite the presence of the hysteresis loop in Fig \eqref{fig:rep_per} tells us that a modulation with period $T= 10^5 t_0$ is not slow enough to be considered adiabatic (at least for $\mathcal{R}<0.01$), the fact that during phase-1 all the curves follow the same path is reassuring: this means that we do not really need to simulate the system for very long times $T$, but that we can use an intermediate value $T\approx 10^4 t_0$ to simulate phase-1 (in fact, no differences between the cases $T=10^4t_0$ and $T=10^5t_0$ have been found during phase-1). Therefore, in the adiabatic limit also the repinning phase should follow the same path.

For this reason we fix $T=2 \times 10^4t_0$ in the following, and we simulate only the first quarter of the hysteresis loop. The results found in this way will be interpreted as the only possible lag-dependence in the limit of very slow lag modulations (at least if the periodic potential is used, the disordered case turns out to be more subtle). In fact, as discussed at the and of Sec \ref{sec_parameters}, lag modulations in a real pulsar are expected to proceed on a timescale that is several orders of magnitude larger than the $T$ values tested, so that 
the hysteresis loop disappears and the adiabatic limit is recovered.




\subsection{Small lag regime: disordered pinscape} 

We have seen in the previous subsection that (in the case of a  periodic potential) the vortices are pinned till $v_{np}(t)$ reaches a well defined critical value for the unpinning $v^*$. For the periodic pinscape in \eqref{forzanew},  the value of $v^*$ can be read from Fig \ref{fig:sinY}, namely  $v^* \approx v_0$.

For the rotational dynamics of pulsars it is important to investigate more closely what happens for small lags  $v_{np}(t) < v^*$, namely when the system is pinned or almost pinned, which corresponds to the non-linear regime sketched in Fig \ref{sketch_MF}. 
This may not sound as an interesting question, since Fig \ref{fig:sinY} clearly shows that $ \langle \dot{x}_2  \rangle = 0 $ for $v_{np}<v_0$. However, perfect pinning may not be realized  when the disordered potential is used.  

The quenched disorder broadens the distribution of pinning forces (as shown in Fig \ref{fig:histogram}), giving rise to a smoothed version of the sharp unpinning threshold observed for the periodic potential at $v^*  \approx v_0$. In fact, during the initialization a significant number of vortices may pin at very weak equilibrium points of the potential, which are absent in the periodic model.
These vortices are also the first that start to move, so that a sharp depinning transition occurring at a well defined $v^*$ seems unlikely for the Gaussian model.
Hence, the effect of disorder in the present context is analogous to the effect of disorder in equilibrium statistical mechanics, where  it is known that microscopic quenched impurities may broaden a sharp first-order phase transition \citep{imry_disorder}.

\begin{figure}
    \centering
    \includegraphics[width=0.47\textwidth]{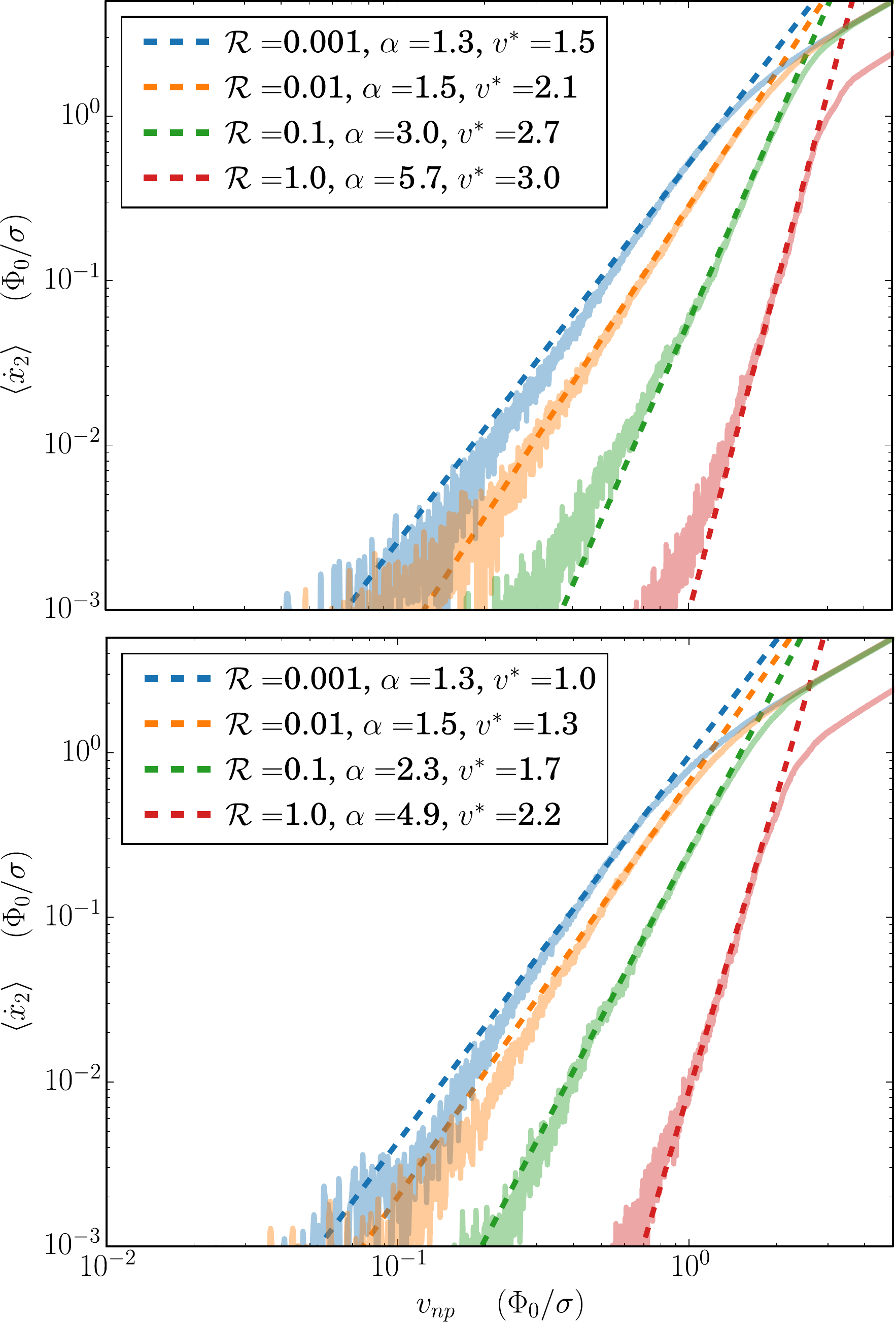}
    \caption{
    The component parallel to the lag  $\langle \dot{x}_2\rangle$ of the average velocity of $N_v=10^4$ vortices  for different values of the drag $\mathcal{R}$. 
        Both $\langle \dot{x}_2\rangle$ and $v_{np}$ are in units of $v_0=\Phi_0/\sigma$.
        The disordered ``attractive'' (upper panel) and ``repulsive'' (lower panel) potentials have been used. The results of the fit \eqref{fitpower} to the data in the region $5\times10^{-3} v_0< \langle \dot{x}_2 \rangle < 0.5 \, v_0$ are reported in the legend (the values of $v^*$ are in units of $v_0$).
    }
    \label{fig:powerlaws}
\end{figure}

To check this behaviour we perform the same kind of simulation described in the previous subsection, but with the disordered potential \eqref{potenziale2D}. 
The results are shown in  Fig \ref{fig:powerlaws}, where we can see that, as expected, the average vortex velocity along the lag $\langle \dot{x}_2\rangle$ is substantially different from zero also in the region of small $v_{np}$.
The form of the curves suggests to assume that $ \gamma_2 =  ( v_{np}/{v^*} )^\alpha $
and to perform a two-parameter fit 
\begin{equation}
    \langle \dot{x}_2  \rangle \, = \, \frac{1}{1+\mathcal{R}^2}v_{np} \left( \frac{v_{np}}{v^*} \right)^\alpha
    \label{fitpower}
\end{equation}
to obtain the values of $v^*$ and $\alpha$. Since for the disordered potential there is no sharp transition between the pinned and unpinned regimes, in this case $v^*$ cannot be interpreted as a critical lag for unpinning. It just tells us that for $v_{np} \lesssim v^*$, the functional form in  \eqref{fitpower} is a valid approximation.

The results of the fit are shown in Fig \ref{fig:powerlaws}. Due to the presence of intrinsic noise discussed in the previous subsection the fit is performed by considering only the  the data satisfying $\langle \dot{x}_2  \rangle > 5\times 10^{-3} v_0$. Furthermore, since we want to remain in the small lag regime, we also impose the upper limit $\langle \dot{x}_2  \rangle < 0.5 v_0$ to the fit region. 
Both the values of $\alpha$ and $v^*$ increase with increasing $\mathcal{R}$ and are not qualitatively different between the ``attractive'' and ``repulsive'' cases.

There is, however, a fundamental question that has to be addressed.
The results in  Fig \ref{fig:powerlaws} have been obtained for $T= 10^5 t_0$ that, as discussed in the previous subsection, produces a modulation of the lag that is slow enough when the periodic potential is used. There is no guarantee that this value of $T$ works also for the disordered potential, i.e. the results of the fit may depend on the value of  $T$ (only in the adiabatic limit $T \rightarrow \infty$ the values of $v_*$ and $\alpha$ are rate-independent).
In fact, performing additional tests with $T= 10^6 t_0$ we find different values of $\alpha$ and $v^*$: in general, as the lag modulation approaches the adiabatic limit, the values of $\alpha$ and $v^*$ increase (e.g. for $\mathcal{R} = 0.1$ we find $\alpha \approx 6$ and $v^* \approx 2.8$, while for $\mathcal{R} = 1$ we get $\alpha \approx 9$ and $v^* \approx 2.9$). This means that in the adiabatic limit we can expect the motion to be entirely suppressed for $v_{np} < v^*$, as in the periodic case. The main difference will be in the shape of the depinning transition, that is smooth in the disordered case.


\section{Comparison of $\gamma_1$ and $\gamma_2$ }
\label{comparisonG1G2}

\begin{figure*}
    \centering
    \includegraphics[width=0.99\textwidth]{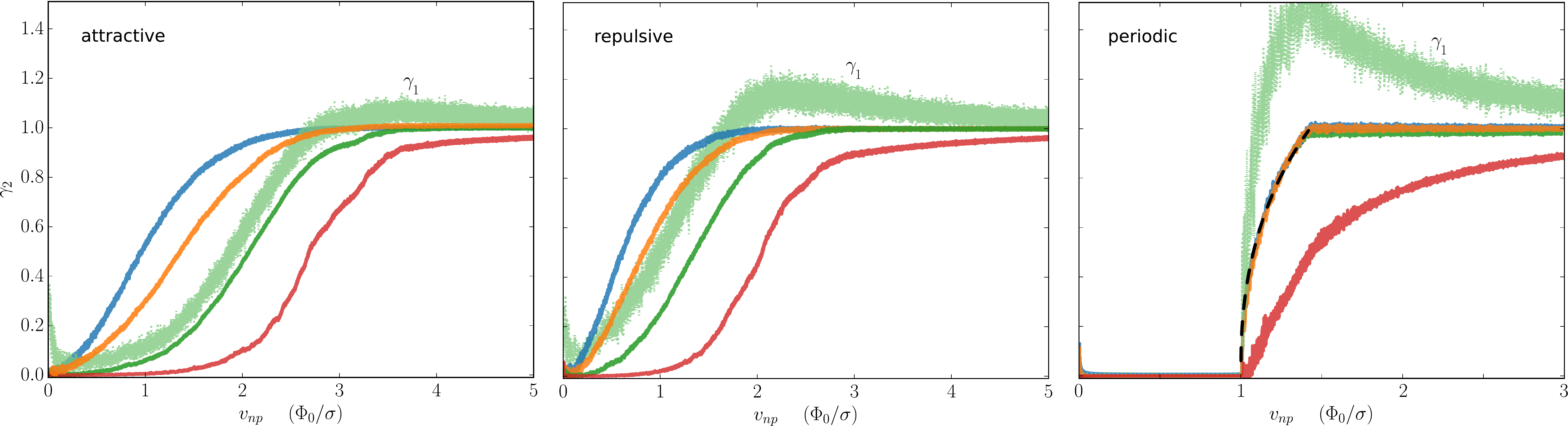}
    \caption{
    The function $\gamma_2(v_{np})$ for the disordered ``attractive'' ($\Phi_0<0$, left), disordered ``repulsive'' ($\Phi_0>0$, center) and periodic (right) potentials with $\sigma = l_P$.
    The same color scheme of Fig \ref{fig:powerlaws} is used: $\mathcal{R}=1$ (red),  $\mathcal{R}=0.1$ (green),  $\mathcal{R}=10^{-2}$ (orange),$\mathcal{R}=10^{-3}$ (blue). 
    The curves in the left panel refer to the curves shown in the upper panel of Fig \ref{fig:powerlaws}.
    For comparison, also the curve $\gamma_1(v_{np})$ is shown, but only for the case  $\mathcal{R}=0.1$ (light-green curve with the $\gamma_1$ label).
    When the periodic potential is used all the curves for $\mathcal{R}\lesssim 1$ are superimposed and, in the range $v_{np} \in [v_0, \sqrt{2} v_0 ]$ are well fitted by the functional form in \eqref{04law}, shown here as a black dashed line. 
    }
    \label{fig:gamma2}
\end{figure*}

\begin{figure}
    \centering
    \includegraphics[width=0.47\textwidth]{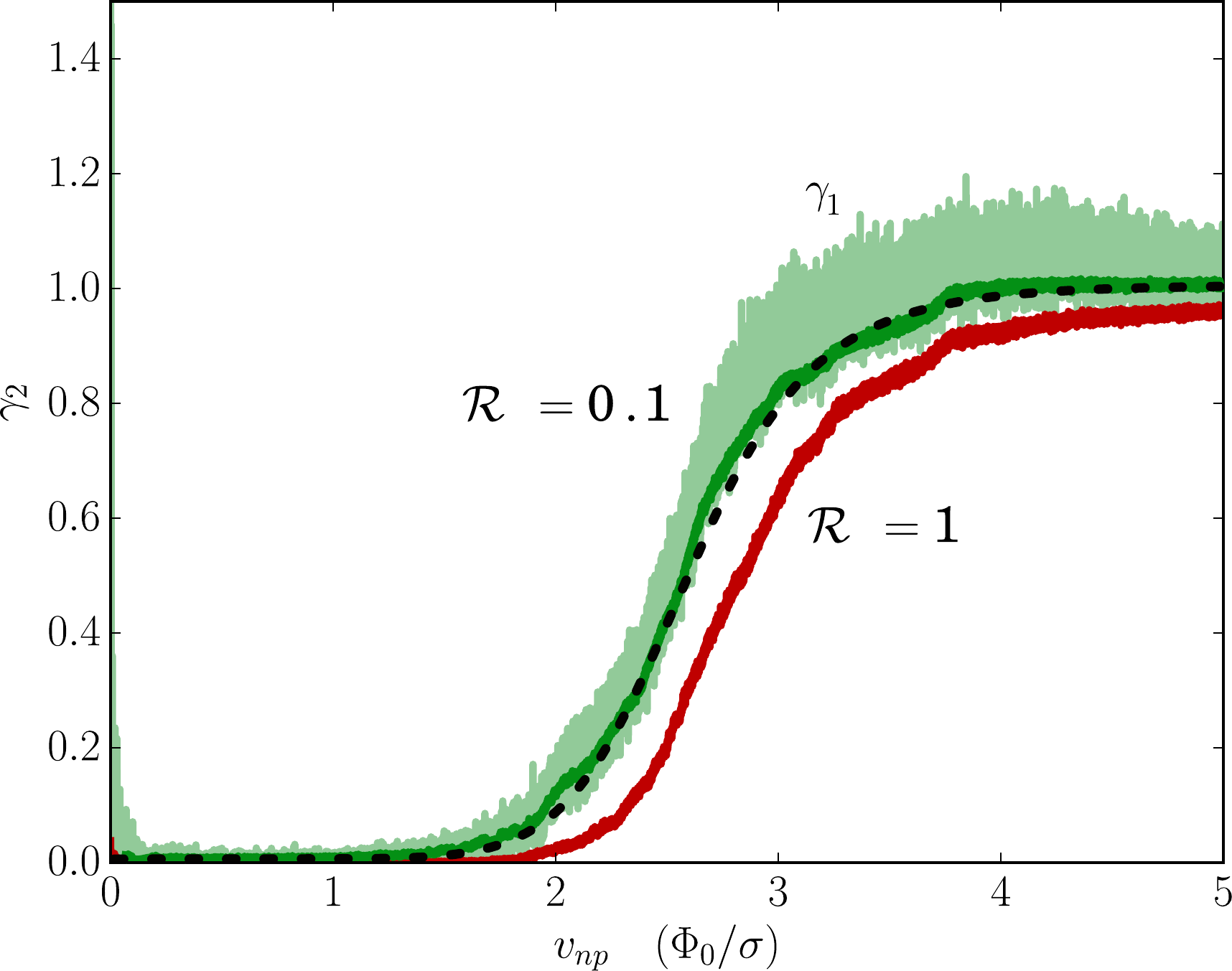}
    \caption{
    The function $\gamma_2(v_{np})$ for the disordered ``attractive'' ($\Phi_0<0$) potential  with $\sigma = l_P$ and $T = 10^6 t_0$.
    The lag  $v_{np}$ is reported in units of $v_0=\Phi_0/\sigma$.
    The same color scheme of Fig \ref{fig:powerlaws} is used: $\mathcal{R}=1$ (red),  $\mathcal{R}=0.1$ (green). 
    For comparison, also the curve $\gamma_1(v_{np})$ is shown, but only for the case  $\mathcal{R}=0.1$ (light-green curve with the $\gamma_1$ label).
    The dotted line corresponds to the fit in equation \eqref{eq_sigmoid} for  $\mathcal{R}=0.1$.
    }
    \label{fig:sigmoid}
\end{figure}

We now ask whether a formal mixture of perfectly pinned and perfectly free vortices is sufficient to reproduce the behaviour of $\langle \dot{\bm{x}}  \rangle$, as discussed in Sec \ref{sec_nonlinear}. 
%
%
%
Therefore, we have to check if the two quantities
\begin{equation}
    \gamma_{1,2} \, = \, \frac{1}{v^{fr}_{1,2} \,  N_v} \sum_{i} \hat{\bm{e}}^i_{1,2} \cdot \langle \dot{\bm{x}} \rangle 
    \label{gamma12i}
\end{equation}
extracted from simulations can be fitted with a single function $\gamma(v_{np})$ such that $0 \leq \gamma \leq 1$. 
The ratio $\gamma_2 $ is shown in Fig \ref{fig:gamma2} for different values of the drag and for both the disordered and periodic potentials (the ``attractive'' and ``repulsive'' disordered cases are obtained with the same data used in Fig \ref{fig:powerlaws}). 
For comparison, the curve $\gamma_1(v_{np}) $ for $\mathcal{R}=0.1 $ is shown. 
This curve is significantly thicker than the ones for $\gamma_2$ because the small denominator $v^{fr}_1$ in \eqref{gamma12i} increases the signal-to-noise ratio when the drag parameter is small.
The condition $\gamma_1 = \gamma_2$ is certainly not met for all the drag parameters and the potentials tested (even though the observed fact that  $\gamma_1 > \gamma_2$ seems more severe for the periodic potential). The observed maximum difference between the two curves is of the order of $40\%$ after the depinning transition, when the periodic potential is used.
For the disordered potential, for both the attractive and repulsive cases, the relative difference between $\gamma_1$ and $\gamma_2$ for small lags is even higher and difficult to quantity precisely.

Again, these observed differences between $\gamma_1$ and $\gamma_2$ could be an artificial effect due to the fact that the evolution is not really adiabatic in our simulations. For this reason we repeat the numerical experiment of Fig \eqref{fig:gamma2} for the disordered attractive potential but with $T=10^6 t_0$. We tested the drags $\mathcal{R}=0.1, \, 1$ and the results are shown in Fig \ref{fig:sigmoid}:  we observe that the differences between $\gamma_2$ and $\gamma_1$ become less pronounced for $\mathcal{R}=0.1$, although still present. The $\gamma_{1}$ curve is not reported for the case $\mathcal{R}=1$, as it is practically superimposed to $\gamma_2$. In particular, the differences between $\gamma_1$ and $\gamma_2$ in the small lag region are much more less pronounced if compared to the ones that can be seen in the first panel of Fig \ref{fig:gamma2}. Both curves can be conveniently fitted with a sigmoid-like function. We assume the form 
\begin{equation}
    \gamma(v_{np}) \, = \, \frac{(v_{np}/v^*)^\alpha}{1+(v_{np}/v^*)^\alpha} 
    \label{eq_sigmoid}
\end{equation}
that is consistent with \eqref{fitpower} in the small lag limit. We find $\alpha =9.1$ 
for both $\mathcal{R}=0.1, \, 1$, but $v^* = 2.5$ when $\mathcal{R}=0.1$ and $v^* = 2.8$ for $\mathcal{R}=1$. 


Finally, it is worth commenting on the periodic case in Fig \ref{fig:gamma2}, which is particularly interesting since it reproduces the theoretical features expected for the unpinning threshold. 
First, even though we are not in the adiabatic limit, all the curves are collapsed on one another for $\mathcal{R}\ll 1$, differently from what happens in the disordered case.
If the evolution were really adiabatic, we could expect the same kind of degeneracy also for the disordered case, on the basis of the universal scaling discussed in the previous section.

We observe  perfect pinning for $v_{np}<v^*$ (in particular, it happens that $v^*=v_0$, see Figs \ref{fig:sinY} and \ref{fig:gamma2}) and a discontinuity in the derivative of $\gamma_2$ at  $v_{np}<\sqrt{2} v_0$. This can be understood in terms of the geometry of the periodic potential. In fact, the two values $v_0$ and $\sqrt{2} v_0$ correspond to the minimum and to the maximum of $|\bm{f}|$ over the square boundaries that separate the stable and unstable regions drawn by the potential $\Phi$. 
At $t=0$ a vortex is pinned at the center of the stable region, but then it start to migrate out as the lag increases, till it reaches the boundary (the smallest value of $v_{np}$ for which the vortex may escape the stable region is the minimum of  $|\bm{f}|$ on the boundary). On the other hand, no vortex can have a bounded motion (which velocity averages to zero) or can find a stable equilibrium position if $v_{np}\geq \max{(|\bm{f}|)} = \sqrt{2} v_0$, so that for $v_{np} > \sqrt{2} v_0$ we are almost in a free-vortex limit (i.e. $\gamma_2=1$). 
In fact, albeit partially hidden by the noise, the change in the derivative of $\gamma_2$ can be seen also for the case $\mathcal{R}=1$ in the right panel of Fig \ref{fig:gamma2}. For the other cases $\mathcal{R}\leq 0.1$ all the curves are superimposed and $\gamma_2$ is well fitted as
\begin{equation}
    \gamma_2 \approx \left(\frac{v_{np}-v^*}{\sqrt{2} v_0 - v^*}\right)^{0.5 \pm 0.05} 
        \label{04law}
\end{equation}
for $\mathcal{R}\ll 1$ and $v^*<v_{np}< \sqrt{2}v_0$, where the critical lag for unpinning is $v^*=v_0$. We remark that this fit is valid for the periodic pinning potential only (see the third panel of Fig \ref{fig:gamma2}) and that the range of values of the phenomenological parameter $v_0$ in a neutron star crust are given in Fig \ref{fig:units}.

\subsection{Summary and discussion of the results }

It may be useful to summarize the main results of our simulations that can be relevant for glitch modelling. 

We extracted the quantities $\gamma_1(v_{np})$ and $\gamma_2(v_{np})$ from simulations, that are directly linked to the mutual friction via equation \eqref{mf_standard}.  
In brief, all the relevant information is contained in Fig \ref{fig:gamma2}. 

First, the fact that the $\mathcal{R}=0.1,0.01,0.001$ curves are not superimposed for the disordered pinscape (i.e. in the first two panels of Fig \ref{fig:gamma2}) has been interpreted as an indication that the lag modulation was not slow enough ($T=10^5 t_0$ ). 
Additional  simulations with $T=10^6 t_0$ show that the  $\mathcal{R}=1$ curve remains almost unchanged (it just shifts a bit on the left), while the one relative to $\mathcal{R}=0.1$ is considerably shifted towards higher lags (so that it is basically superimposed to the red curve in Fig \ref{fig:gamma2}). 
This can be seen by comparing Fig \ref{fig:sigmoid} with the first panel of Fig \ref{fig:gamma2}.
The general trend is that, by increasing $T$, the differences between $\gamma_2$ and $\gamma_1$ become less pronounced, although still present.
This is an indication that the depinning transition for a disordered pinscape can be modelled in the adiabatic limit by assuming that the curves  $\gamma_1\approx \gamma_2$ are smooth sigmoid functions $\gamma(v_{np})$, as in \eqref{eq_sigmoid}. 
Let us remark that the power-law form of  $\eqref{eq_sigmoid}$  has been chosen for consistency with \eqref{fitpower}, but  a more general behaviour has to be expected. 
The exact form of $\gamma(v_{np})$ depends on the details of the pinscape. 

On the other hand, when the periodic potential is used, we find no indication that the differences between $\gamma_2$ and $\gamma_1$ become less pronounced when $T$ is increased: for $T=10^6t_0$ the results are very similar to the ones in the third panel of Fig \ref{fig:gamma2}, in particular the pronounced overshoot of $\gamma_1$ seems to be a stable feature.
In this case $\gamma_1(v_{np})$ and $\gamma_2(v_{np})$ are not smooth sigmoid  functions but have a derivative discontinuity at the depinning point $v_{np} = v^* $ and at $v_{np} = \sqrt{2} v_0$. In this interval of lags, the behaviour of  $\gamma_2(v_{np})$ is well described by equation \eqref{04law}.


\section{Conclusions}
\label{sec:conclusions}

We developed a kinetic approach to extract the vortex-mediated mutual friction as a function of the velocity lag between the normal and superfluid components in a neutron star.  
This approach is in principle applicable also to other systems, like laboratory superfluids and type-II superconductors and is, in fact, based on the same kind of many-vortices simulations used to investigate flux-tube dynamics in superconducting samples \citep{Mawatari1997PhRvB}.

As a first step, we considered a minimal model able to reproduce the non-linear features of mutual friction expected in real neutron stars, namely the fact that there should be a critical lag $v^*$  that separates two regimes: a sub-threshold regime where the dissipation induced by drag forces is suppressed and a super-critical regime in which the mutual friction recovers the linear HVBK form.
Our model is minimal in the sense that additional non-linearities due to an explicit dependence of the drag parameter on the vortex velocity \citep{celora2020MNRAS} have not been considered. 
Also vortex tension \citep{link2009PhRvL,Haskellhop} and mutual vortex interactions \citep{war_mel_2013,fily2010PhRvB} have been neglected.

In this minimal setting the presence of a depinning transition is a pure effect of the competition between the pinning potential  and the Magnus force. 

In a periodic potential we found a sharp depinning transition occurring at a well defined critical lag, which value  $v^*$ depends on the parameters that define the pinning landscape.
When such a periodic pinning potential is used, our simulations indicate that the effect of having a more or less effective dissipation (set by the drag parameter $\mathcal{R}$) does not change the critical value $v^*$, at least  in the adiabatic limit.
On the other hand, if the non-homogeneous medium is highly disordered, resulting in a broader distribution of possible local pinning forces, the depinning transition turns out to be smooth and a well defined critical lag $v^*$ does not exist anymore. 

In this sense, the disorder plays a role similar to that of temperature in vortex-creep models \citep{Alpar84a}, as the temperature is expected to smooth the depinning transition. 
This may result (at least in the limit considered here, where interactions between vortices are neglected) into a more continuous release of angular momentum via more frequent but smaller glitches \citep{mcKenna_Nat_90,montoli_eos_2020}. 
This leads us to speculate that in a periodic potential the sharp depinning transition may lead to a narrower distribution of glitch sizes, while in a disordered potential the transition is not well defined and thus glitch sizes are likely to be  more broadly distributed.  
Hence, the different glitching behaviour of pulsars may also be due to glitches originating in different regions of the star with different pinning regimes (e.g. if vortices pin in the pasta phase layer at the bottom of the inner crust this would constitute a realization of interaction with a disordered pinscape).

Another qualitative point is that the level of dissipation sets the timescale with which the system responds to an externally imposed modulation of the lag: a fast modulation can result in a rate-dependent hysteresis of the mutual friction, especially if the system is weakly dissipative. 
This immediately implies that, for a fast modulation of the background lag, the threshold for repinning cannot be greater than  the one for unpinning, at least in our model where the interactions between vortex lines are not taken into account. 
It would be interesting to check if the presence of interactions  between vortex lines would promote the hysteresis to a rate-independent one \citep{fily2010PhRvB}. This would provide an automatic load and discharge mechanism for pulsar glitch models, where vortices unpin close to a certain value of the lag but must move till the lag relaxes to a smaller value before having a chance to repin.

From the quantitative point of view we were able to provide a functional form for the ratio $\gamma_2$ that defines the non-linear behavior of the mutual friction component orthogonal to the lag, see \eqref{mf_standard}. For the periodic potential, the form of $\gamma_2$ is the one in equation \eqref{04law}, while for the disordered model we were only able to assess that the depinning transition is smooth and that $\gamma_2\approx \gamma_1$ have the form of a sigmoid  function. 

There is also a theoretical issue that deserves further investigation, namely the question to which extent it is possible to assume that $\gamma_1\approx\gamma_2$. Our preliminary simulations show that this could be the case in the adiabatic limit of very slow lag variations, which is the relevant limit for pulsar glitch modelling. 
Apart from being an interesting theoretical question, this is also of considerable practical value as it is much easier to resolve the average motion of the vortices in the direction parallel to the lag.  We plan to investigate this in a forthcoming work.

Finally, let us remark that our approach is purely phenomenological and that the link with the internal physics of neutron stars is provided by the interpretation of the hydrodynamic variables in the system \eqref{pizzobaldo} and by the 
values of the physical units of length $\sigma$, velocity $v_0$ and time $t_0=\sigma/v_0$ given in Fig \ref{fig:units}. 
Clearly, the present approach is applicable also to the case of superfluid $^4$He: a laboratory realization of the system studied here would be an Helium film over a substrate that could provide enough roughness to pin vortices, namely a two-dimensional version of the series of experiments reported by \cite{Tsakadze1980}.

\section*{Acknowledgements}

Partial support comes from PHAROS, COST Action CA16214. Marco Antonelli acknowledges support from the Polish National Science Centre grant SONATA BIS 2015/18/E/ST9/00577, P.I.: B. Haskell. 

\section*{DATA AVAILABILITY}

No new data were generated or analysed in support of this research. 
Simulations output and codes will be shared upon reasonable request.


\appendix

\section{General solution of the linear equations of motion for straight vortices}
\label{app_solution}

The classic  derivation of the mutual friction force in the case of superfluid $^4$He \citep{Hall1956_II,Bekarevich1961} and neutron star interiors \citep{mendell1991_II,langlois98,AndSid06} is extended to situations in which there is also a generic external field $\langle \bm{f} \rangle $ that acts on the vortices.
An alternative extension of the classical approach to include the effect of pinning with flux-tubes in the core is presented in \cite{sourie2020_pinningCore}.

From the phenomenological point of view, the most general equation of motion for a straight vortex segment (i.e. a piece of vortex with no internal dynamics) immersed into two distinct background flows can be written on the basis of purely geometric considerations. We represent a vortex segment as a point particle, meaning that all the geometric degrees of freedom of an extended line that can bend are frozen in this description. 
Furthermore, imposing $\hat{\bm{\kappa}} = \langle \hat{\bm{\kappa}} \rangle $, implies that the lines are locally parallel (they are organized in parallel bundles that can bend only on length-scales bigger than the extension of the macroscopic fluid element considered). 

We indicate with $\bm{v}_n$ and $\bm{v}_p$ the two background velocity fields and with $\bm{v}_L$ the vortex segment velocity in a generic frame.
In the overdamped regime, the most general equation of motion for the vortex segment that is up to the first order in the relative velocities $\bm{v}_L-\bm{v}_n$ and $\bm{v}_L-\bm{v}_p$ is 
\begin{equation}
{M}^n \, (\bm{v}_L-\bm{v}_n) + {M}^p \, (\bm{v}_L-\bm{v}_p) \, + \langle \bm{f} \rangle  = \, 0 \, ,
\label{eq:ftot}
\end{equation}
where the two  matrices ${M}^{\rm{x}}$ are functions of $\hat{\kappa}$ and of some microphysical parameters that tune the interaction between the vortex line and the background flows.
The  ${M}^{\rm{x}}$ can be constructed as a linear combination of three fundamental operators, $K$, $\perp$ and $\parallel$,
\begin{equation}
\begin{split}
{M}^{n}_{ ij} &=  K_{ij}  - \, \xi_\perp \, \perp_{ij}  - \, \xi_\parallel \, \parallel_{ij}  
\\
{M}^{p}_{ ij} &= \mathcal{R}_\times  \, K_{ij}  - \, \mathcal{R}_\perp \perp_{ij}  
- \, \mathcal{R}_\parallel \, \parallel_{ij} \, ,
\end{split}
\end{equation}
where
\begin{equation}
\begin{split}
K_{ij} & =  \epsilon_{iaj} \hat{\kappa}^a  
\\
\perp_{ij} & =  \delta_{ij} - \hat{\kappa}_i \hat{\kappa}_j = -K_{ia}K_{aj}
\\
\parallel_{ij} & =   \hat{\kappa}_i \hat{\kappa}_j     \, .
\end{split}
\end{equation}
The terms proportional to the dimensionless coefficients $\xi_\parallel $, $\xi_\perp$, $\mathcal{R}_\parallel$ and   $\mathcal{R}_\perp$ have an explicit minus sign since they arise from some kind of friction between the vortex and the currents. On the contrary, we do not specify the sign of $\mathcal{R}_\times$ since it plays the role of a charge parameter in a Lorentz-like force, which may be positive or negative. From the microscopic point of view, the calculation of these coefficients is a difficult and subtle task, which should be done consistently with the physical meaning attributed to the fields $\bm{v}_n$ and~$\bm{v}_p$.
 
It is convenient to work in the frame of the p-fluid: in this frame we indicate the vortex velocity and the velocity of the n-fluid as $\langle \dot{\bm{x}} \rangle =\bm{v}_{L}-\bm{v}_{p}$  and $\bm{v}_{np}=\bm{v}_{n}-\bm{v}_{p}$ respectively. 
Solving \eqref{eq:ftot} for the vortex velocity gives
\begin{equation}
\begin{split}
\langle \dot{\bm{x}} \rangle  \, = \, & ({M}^{n}+{M}^{p})^{-1} \, ( M^n \bm{v}_{np} -  \langle \bm{f} \rangle )
\\
 \, = \, & (B_\perp \perp  + B_\parallel \parallel  - B_\times \, K) \bm{v}_{np} + \\
 &  \, + \, (C_\perp \perp  + C_\parallel \parallel  +  C_\times \, K) \langle \bm{f} \rangle 
\label{eq:sol1} 
\end{split}
\end{equation}
where
\begin{equation}
\begin{split}
B_\perp  &= ( 1 +\mathcal{R}_\times + \xi_\perp \mathcal{R}_\perp + \mathcal{R}_\perp^2 ) /D
\\
B_\times   &= (  \mathcal{R}_\perp - \mathcal{R}_\times  \xi_\perp  ) /D
\\
C_\perp &= ( \mathcal{R}_\perp + \xi_\perp ) /D
\\
C_\times &=  ( 1+  \mathcal{R}_\times    ) /D
\\
D &= (1+\mathcal{R}_\times)^2 +  ( \xi_\perp + \mathcal{R}_\perp )^2
\end{split}
\end{equation}
and
\begin{equation}
\begin{split}
B_\parallel & =   \xi_\parallel /( \xi_\parallel  + \mathcal{R}_\parallel ) 
\\
C_\parallel & = 1/( \mathcal{R}_\parallel + \xi_\parallel ) 
\end{split}
\end{equation}
In neutron star interiors the parameters $\xi_\parallel$ and $\xi_\perp$ are taken to be zero \citep[see e.g.][]{Carter_Prix_Magnus,sourie2020_pinningCore}, so that the mutual friction force is proportional to the Magnus force, in accordance with \eqref{fmf_magnus}. 
Remembering that $\perp = -K^2$ and $K\perp = K$, it is immediate to find
\begin{equation}
\dfrac{\bm{F}_{n}}{|\boldsymbol{\omega}_n|} = 
\mathcal{B}_\times K \bm{v}_{np} 
-
\mathcal{B}_\bot \bot \bm{v}_{np}
-
\mathcal{C}_\times K \langle \bm{f} \rangle
+
\mathcal{C}_\bot \bot  \langle \bm{f} \rangle
\, ,
\label{fmf_totttale}
\end{equation}
that is equivalent to the more familiar HVBK-like form
\begin{multline}
\bm{F}_{n}= 
\mathcal{B}_\times  \boldsymbol{\omega}_n \times \bm{v}_{np} 
+
\mathcal{B}_\bot \hat{\boldsymbol{\omega}}_n \times(\boldsymbol{\omega}_n \times \bm{v}_{np}) -
\\
-
\mathcal{C}_\times \boldsymbol{\omega}_n \times \langle \bm{f} \rangle
-
\mathcal{C}_\bot  \hat{\boldsymbol{\omega}}_n \times(\boldsymbol{\omega}_n \times  \langle \bm{f} \rangle )
\, .
\label{fmf_tottta}
\end{multline}
The signs in \eqref{fmf_totttale} and \eqref{fmf_tottta} have been chosen so that the coefficients  
\begin{equation}
\begin{split}
\mathcal{B}_\times  &= 1 - {B}_\bot 
\\
\mathcal{B}_\bot  &= {B}_\times 
\\
\mathcal{C}_\times &= C_\bot
\\
\mathcal{C}_\bot &=  C_\times
\end{split}
\end{equation}
are all positive. Note that the coefficients $ \mathcal{R}_\parallel $ and $ \xi_\parallel $ do not contribute to the final form of the mutual friction: their role is just to guarantee that an inverse of the $3\times 3$ matrices $M^p$ and $M^n$ exists. Alternatively, $\mathcal{R}_\parallel $ and $ \xi_\parallel $ could be dropped 
altogether by restricting the inverse definition to be $(M^{\rm{x}})^{-1} M^{\rm{x}} = M^{\rm{x}}(M^{\rm{x}})^{-1} = \bot $.

Finally, thanks to a change of chemical basis \citep{CarterKhalat92,thermo2020CQG}, it is possible to translate the above results into the more common formalism usually employed in the study of He-II, see e.g. section IV-D of \cite{prix2004} or sections 4.2 and 4.7 of \cite{thermo2020CQG} for the relativistic analogue.


\section{Correlation functions}
\label{app_corr}

Consider a potential in $D$ spatial dimensions of the form
\begin{equation}
\label{potenzialeD}
 \Phi_{N_P}(\bm{x}) =  \sum_a \Phi_1 (|\bm{x}-\bm{r}_a|)  -  c_{N_P} \, , 
\end{equation}
where the $\bm{r}_a$ are $N_P$ independent random variables, identically distributed with law $p(\bm{r}_{a})$ over the volume $[-L/2,L/2]^D$.
The constant term $ c_{N_P}$ is non-physical and can be added to ensure that $\Phi_{N_p}(\bm{x})$ has zero average.
For a function $g$ of the positions $\bm{r}_a$ and $\bm{x}$, 
we define
\begin{equation}
\label{average_g}
 \langle \, g (\bm{x})\, \rangle_{N_P} 
 =
 \int \prod_{a} d^{D}\!r_a \, p(\bm{r}_{a}) \, g(\bm{x} ; \bm{r}_1,...,\bm{r}_{N_P}) \, .
\end{equation}
The limit of large $N_P$ and $L$ is always taken in a way that $n_P = N_P/L^D$ is finite (we drop the subscript $N_P$ when this limit is taken).
If the law $p$ and $\Phi_1$ are such that the self-averaging property 
 \begin{equation}
 \label{auto_av}
   \langle \, \Phi_{N_P}(\bm{x}) \, \rangle_{N_P} \, 
   \rightarrow  \langle \, \Phi(\bm{x}) \, \rangle
   = \, n_P \int d^{D}r \, 
   \Phi_1(\bm{x}-\bm{r} ) \, 
 \end{equation}
holds (as in the case of Sec. \ref{disordinatoo}, where the single contribution $\Phi_1$ is Gaussian), it is possible to show that
\begin{equation}
\label{average_phiNp}
c_{N_P} \,  
 \rightarrow  c    = \, n_P \int d^{D}\!r \,  \Phi_1(|\bm{r}|)  \, .
\end{equation}
The large $N_P$ limit of the correlation function
\begin{equation}
\label{corrNp}
C_{N_P}( \bm{x},\bm{y} ) 
\,= \,  
\langle \,    \Phi_{N_P}(\bm{x})   \Phi_{N_P}(\bm{y})   \, \rangle_{N_P}   \, 
\end{equation}
must depend only on the norm $|\bm{x}-\bm{y}|$: by using \eqref{potenzialeD} and \eqref{average_g}, a direct calculation gives 
\begin{equation}
\label{corr}
C( |\bm{z}| ) 
=
\langle \,    \Phi(\bm{z})   \Phi(0)   \, \rangle  =
n_P \! \int d^D\!r \, \Phi_1(|\bm{z}-\bm{r}|) \Phi_1(|\bm{r}| ) \, . 
\end{equation}
Note that, since $\Phi_{N_P}$ is a sum of independent and identically distributed random variables $\Phi_1$, the above results are consistent with the usual formulation of the Central Limit Theorem. 

In the present work the potential \eqref{potenzialeD} is a convenient tool to construct a disordered force, which is the physical field entering into the equation of motion for the vortex,
\begin{equation}
\label{fD}
\bm{f}_{N_P}(\bm{x})  = -\nabla   \Phi_{N_P}(\bm{x})  
 = 
 \sum_a \bm{f}_1 ( \bm{x}-\bm{r}_a )  \, , 
\end{equation}
where 
\begin{equation}
\label{f1}
 \bm{f}_{1}(\bm{x}) 
 = 
\frac{\bm{x}}{|\bm{x}|}  \Phi_1'(|\bm{x}|)  \, . 
\end{equation}
Considering the force as arising from a potential is not strictly needed, but it simplifies the calculation of the correlation function 
\begin{equation}
\label{corrFNp}
D^{ij}_{N_P}( \bm{x},\bm{y} ) 
\,= \,  
\langle \,   f^i_{N_P}(\bm{x})   f^j_{N_P}(\bm{y})   \, \rangle_{N_P}   \, 
\end{equation}
and it guarantees that 
\begin{equation}
  \langle \,   \bm{f}_{N_P}(\bm{x})     \, \rangle_{N_P} 
  \rightarrow
  \langle \, \bm{f}(\bm{x}) \, \rangle \, = \,  0 
\end{equation}
in the limit of large $N_P$. 
In fact, instead of carrying out a direct calculation, the correlation is more easily found by 
\begin{equation}
\label{corrFNp}
D^{ij}_{N_P}( \bm{x},\bm{y} ) 
\,= \,  
 \frac{\partial^2}{\partial {x^i} \, \partial{y^j}}  C_{N_P}( \bm{x},\bm{y} )  \, .
\end{equation}
In the large $N_P$ limit the correlation is expected to be translation invariant, so that 
(note the extra minus with respect to the above equation)
\begin{equation}
\label{corrF}
D_{ij} ( \bm{z} ) 
\,= \,  
 - \frac{\partial^2}{\partial {z^i} \, \partial{z^j}}  C( |\bm{z}| ) 
\,= \,  
- \bot_{ij}  \frac{C'}{|\bm{z}|} - \parallel_{ij} C''
 \, ,
\end{equation}
where the projectors $ \bot_{ij}$ and $ \parallel_{ij}$ are defined as
\begin{equation}
    \label{projectors}
   \bot_{ij} \, = \, \delta_{ij} - \frac{{z^i} \, {z^j}}{|\bm{z}|^2}
 \qquad \quad
  \parallel_{ij} \, = \, \frac{{z^i} \, {z^j}}{|\bm{z}|^2} \, .
\end{equation}
It may be interesting to consider the trace of \eqref{corrF}: since the trace of $\bot$ and $\parallel$
is $D-1$ and 1 respectively, 
\begin{equation}
\label{traceF}
\langle \, \bm{f}(\bm{z})  \cdot \bm{f}(0)   \, \rangle 
\,= \,  
(1-D) \frac{C'}{|\bm{z}|} - C''
 \, .
\end{equation}
The $|\bm{z}|\rightarrow 0 $ limit of expressions like  \eqref{corrF} or \eqref{traceF} is more easily calculated by means of an equivalent formula for the correlation: for $\bm{z} = \bm{x}-\bm{z}$ we have 
\begin{multline}
\label{corrFNp2}
D_{ij} ( \bm{x}-\bm{y} ) 
 \,= \,  
 \frac{\partial^2}{\partial {x^i} \, \partial{y^j}}  C( |\bm{x}-\bm{y}| )  \,  
 \\
 \,= \,  
n_P \int d^D\!r \, \Phi_1'(|\bm{r}-\bm{z}|) \Phi_1'(|\bm{r}|) \frac{r^i-z^i}{|\bm{r}-\bm{z}|} \frac{r^j}{|\bm{r}|} \, ,
\end{multline}
that immediately gives 
\begin{align}
\label{corrFNp0}
D_{ij} ( 0 ) 
\,= \,  
n_P \int d^D\!r \, \Phi_1'(|\bm{r}|)^2 \frac{r^i r^j}{|\bm{r}|^2}  \, 
\,= \,  \frac{n_P \delta_{ij} }{D} \int d^D\!r \, \Phi_1'(|\bm{r}|)^2
\end{align}
and 
\begin{equation}
\label{traceF0}
\langle \, \bm{f}(0)  \cdot \bm{f}(0)   \, \rangle 
\,= \,  
  n_P  \int d^D\!r \, \Phi_1'(|\bm{r}|)^2
 \, .
\end{equation}
Using a Gaussian form for $\Phi_1$  as in \eqref{potenziale2D} and random variables $\bm{r}_a$ that are distributed uniformly according to $p(\bm{r}_a)=L^{-D}$, it is possible to obtain \eqref{gaussianC} for  $D=2$. 
More generally, the correlation and the constant $c$ read
\begin{align}
\label{corrD}
& C( |\bm{z}| ) 
   =
\Phi_0^2 \, n_P  \, \sigma^D  \, \pi^{D/2}  \, e^{-|\bm{z}|^2/4\sigma^2}  
\\
\label{cD}
& c \, = \, \Phi_0 \, n_P  \, \sigma^D  \, (2 \pi)^{D/2}   \,  . 
\end{align}
Similarly, the correlation function that generalises \eqref{pistacchio} to $D$ spatial dimensions is
\begin{equation}
\label{corrDf}
 D_{ij}( |\bm{z}| ) 
   =
\Phi_0^2 \, n_P  \, \sigma^{D-2}  \, \pi^{D/2} 
\left( \frac{\delta_{ij} }{2}-\frac{|\bm{z}|^2 }{4\sigma^2}  \parallel_{ij}  \right) 
e^{ -\frac{|\bm{z}|^2}{4\sigma^2} }  \, ,
\end{equation}
while the variance of the random variable $\bm{f}(\bm{x})$ is 
\begin{equation}
\label{traceF0gauss}
\langle \, \bm{f}(0)  \cdot \bm{f}(0)   \, \rangle 
=
\frac{D}{2} \Phi_0^2 \, n_P  \, \sigma^{D-2}  \, \pi^{D/2} 
 \, .
\end{equation}
The above relations can be useful to extend the present treatment to the case $D=3$.


\bibliographystyle{mnras}
\bibliography{biblio}

\bsp
\label{lastpage}
\end{document}